\documentclass[aps,prl,twocolumn]{revtex4-2}
\usepackage{graphicx,color}
\usepackage{dcolumn}
\usepackage{amssymb,amsmath,amsbsy,bm,amscd}
\usepackage[colorlinks=true,citecolor=blue]{hyperref}
\usepackage[table]{xcolor}
\usepackage{physics}
\usepackage{tabularx,ragged2e}

\begin{document}

\title{Statistical Analysis on Random Quantum Sampling by Sycamore and Zuchongzhi Quantum Processors}
\author{Sangchul Oh}
\email{oh.sangchul@gmail.com}
\affiliation{Department of Chemistry, Department of Physics and Astronomy, and Purdue Quantum Science and
    Engineering Institute, Purdue University, West Lafayette, IN, USA}
\author{Sabre Kais}
\email{Corresponding author: kais@purdue.edu}
\affiliation{Department of Chemistry, Department of Physics and Astronomy, and Purdue Quantum Science and
    Engineering Institute, Purdue University, West Lafayette, IN, USA}
\date{\today}

\date{\today}

\begin{abstract}
Random quantum sampling, a task to sample bit-strings from a random quantum circuit, is considered one of 
suitable benchmark tasks to demonstrate the outperformance of quantum computers even with noisy qubits. 
Recently, random quantum sampling was performed on the Sycamore quantum processor with 53 qubits 
[Nature {\bf 574}, 505 (2019)] and on the Zuchongzhi quantum processor with 56 qubits 
[Phys. Rev. Lett. {\bf 127}, 180501 (2021)]. Here, we analyze and compare statistical 
properties of the outputs of random quantum sampling by Sycamore and Zuchongzhi. 
Using the Marchenko-Pastur law and the Wasssertein distances, we find that quantum random sampling of 
Zuchongzhi is more closer to classical uniform random sampling than those of Sycamore. 
Some Zuchongzhi's bit-strings pass the random number tests while both Sycamore and Zuchongzhi show similar 
patterns in heatmaps of bit-strings. It is shown that statistical properties of both random quantum samples 
change little as the depth of random quantum circuits increases. Our findings raise a question about 
computational reliability of noisy quantum processors that could produce statistically different outputs 
for the same random quantum sampling task.
\end{abstract}
\maketitle

\paragraph*{\it Introduction.---}
A quantum computer is believed to simulate quantum systems much better~\cite{Feynman1982} and to 
solve some computational tasks exponentially faster~\cite{Shor1997,HHL2009} than a classical computer.  
A demonstration of the outperformance of a quantum computer, called quantum 
supremacy~\cite{Preskill2012} or quantum advantage, is considered one of the important milestones in 
developing practical quantum computers. Random quantum sampling~\cite{Aaronson2013} 
is regarded as a good candidate for demonstrating quantum advantage with noisy intermediate-scale quantum 
(NISQ) computers available these days. Recently, quantum advantage has been claimed for random 
quantum sampling on the Sycamore quantum processor with 53 superconducting qubits~\cite{Arute2019} 
and the Zuchongzhi quantum processor with 56 superconducting qubits~\cite{Wu2021}, and for Boson 
sampling with optical qubits~\cite{Aaronson2013,Zhong2020,Zhong2021}.

Quantum advantages of the Sycamore and Zuchongzhi quantum processors over classical computers for random 
quantum sampling were verified using the linear cross-entropy benchmarking (XEB) fidelity, whose 
values were estimated slightly larger than zero~\cite{Arute2019,Wu2021}. Both quantum processors 
are made of two-dimensional arrays of superconducting transmon qubits, have similar error rates, 
executed the same random quantum circuit, and scored similar XEB values. So, one may speculate 
that the output bit-strings generated by the two noisy quantum processors would be statistically 
close. However, it is unknown whether two noisy quantum processors with similar values of the XEB 
fidelity sample statistically similar bit-strings or not. In this paper, we analyze the statistical 
closeness of the outputs of the two noisy quantum processors using the NIST random 
number tests~\cite{NIST2010}, the Marchenko-Pastur distribution of eigenvalues of random matrices 
of bit-strings~\cite{MarPas67}, and the Wasserstein distance between samples~\cite{Villani2008}. 
We show that Sycamore's random bit-strings are farther away from classical uniform random 
bit-strings than Zuchongzhi's outputs, that is, the two outputs of bit-strings are noticeably 
different. 

\paragraph*{Random quantum sampling.---}
Let us first summarize random quantum sampling implemented on the Sycamore 
and Zuchongzhi quantum processors~\cite{Arute2019,Wu2021}.
The task of random quantum sampling is to sample bit-strings $x = a_1\cdots a_n \in \{0,1\}^n$ 
by applying a random quantum circuit $U$ on the initial state $\ket{0}$ of $n$ qubits 
followed by the measurement. Both Sycamore and Zuchongzhi quantum processors executed the same 
random quantum circuit $U$ composed of $m$ cycles as follows. A quantum circuit $U_k$ at the $k$-th 
cycle consists of single-qubit gates $R$ selected randomly from the set $\{\sqrt{X},\sqrt{Y},\sqrt{W}\}$ on all qubits 
and deterministic two-qubit gates on the pair of qubits selected in the sequence of the coupler 
activation patterns of 2-dimensional superconducting qubits. 
After $m$ cycles, 
a final single-qubit gate $R$ is applied before the measurement, so the whole random quantum circuit
with cycle $m$ is given by $U = RU_m U_{m-1}\cdots U_1$.
The output of the random quantum circuit is a bit-string $x$ that is sampled from the probability 
$p(x) = |\bra{x}U\ket{0}|^2$. By implementing the same random quantum circuit $M$ times, 
a collection of $M$ random bit-strings ${\cal D} = \{x_1,\dots,x_{M}\}$, i.e., 
an $M\times n$ binary bit array, is obtained. 

\begin{figure*}[t]
\includegraphics[width=0.32\textwidth]{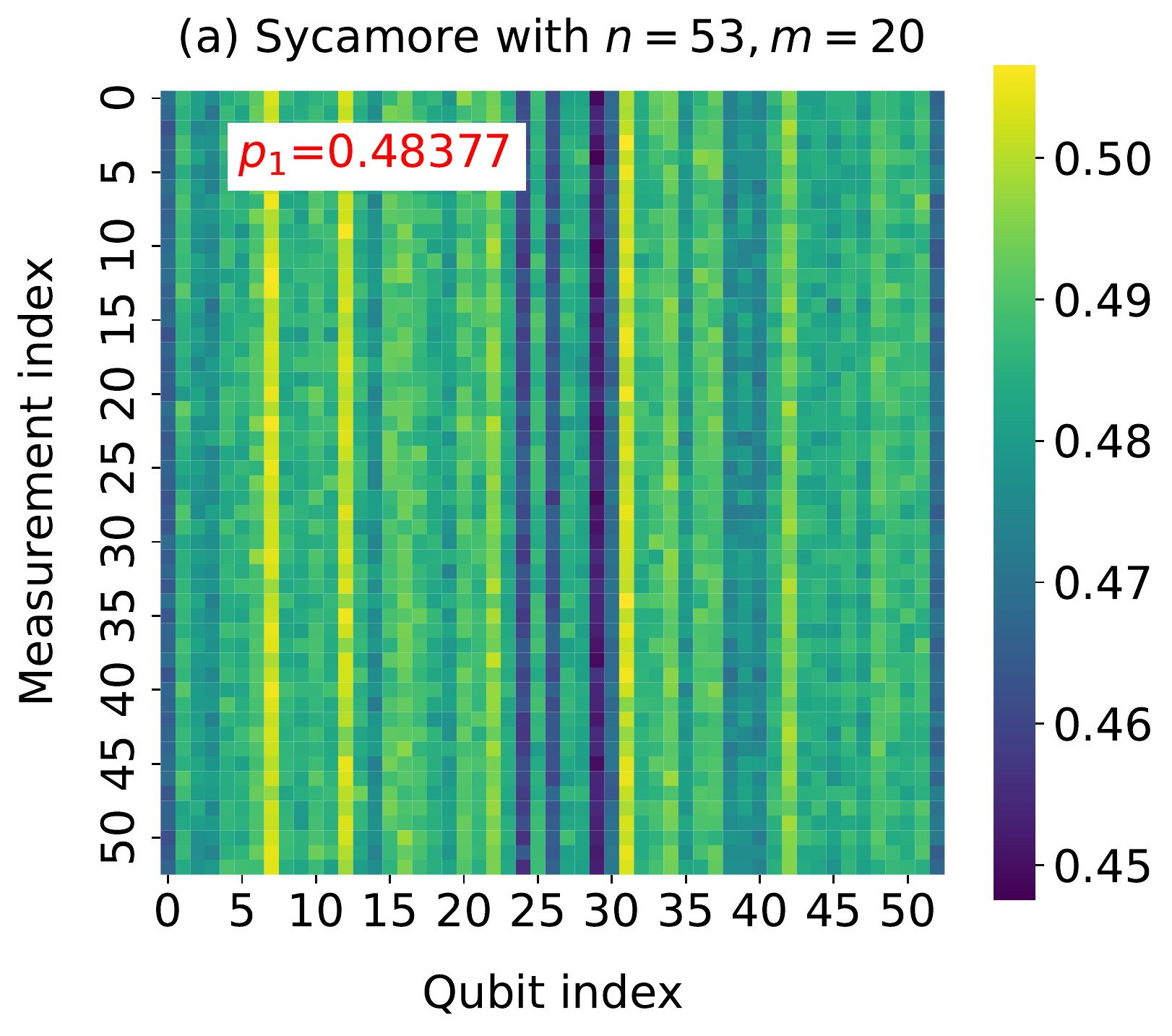}\hfill
\includegraphics[width=0.32\textwidth]{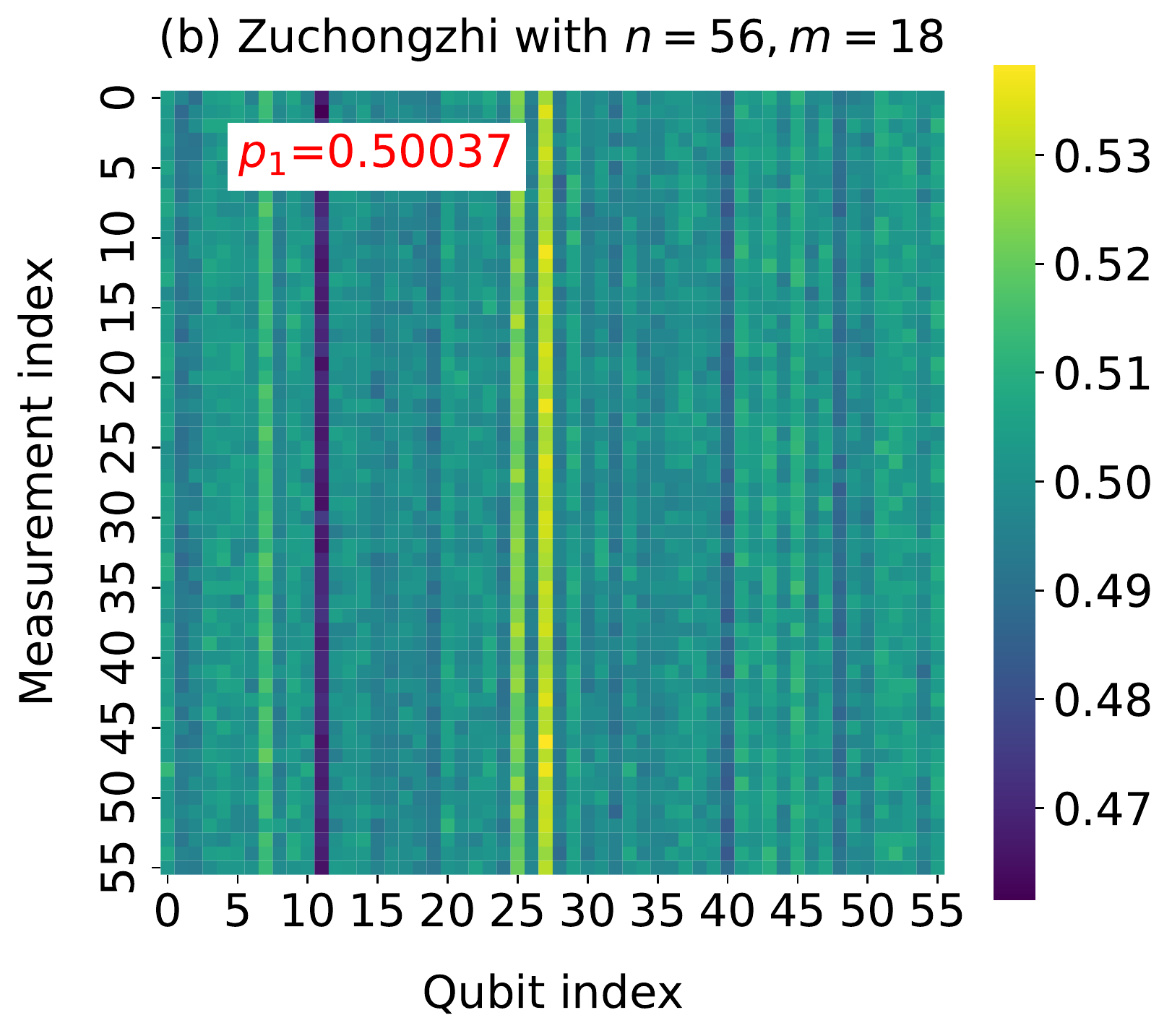}\hfill
\includegraphics[width=0.32\textwidth]{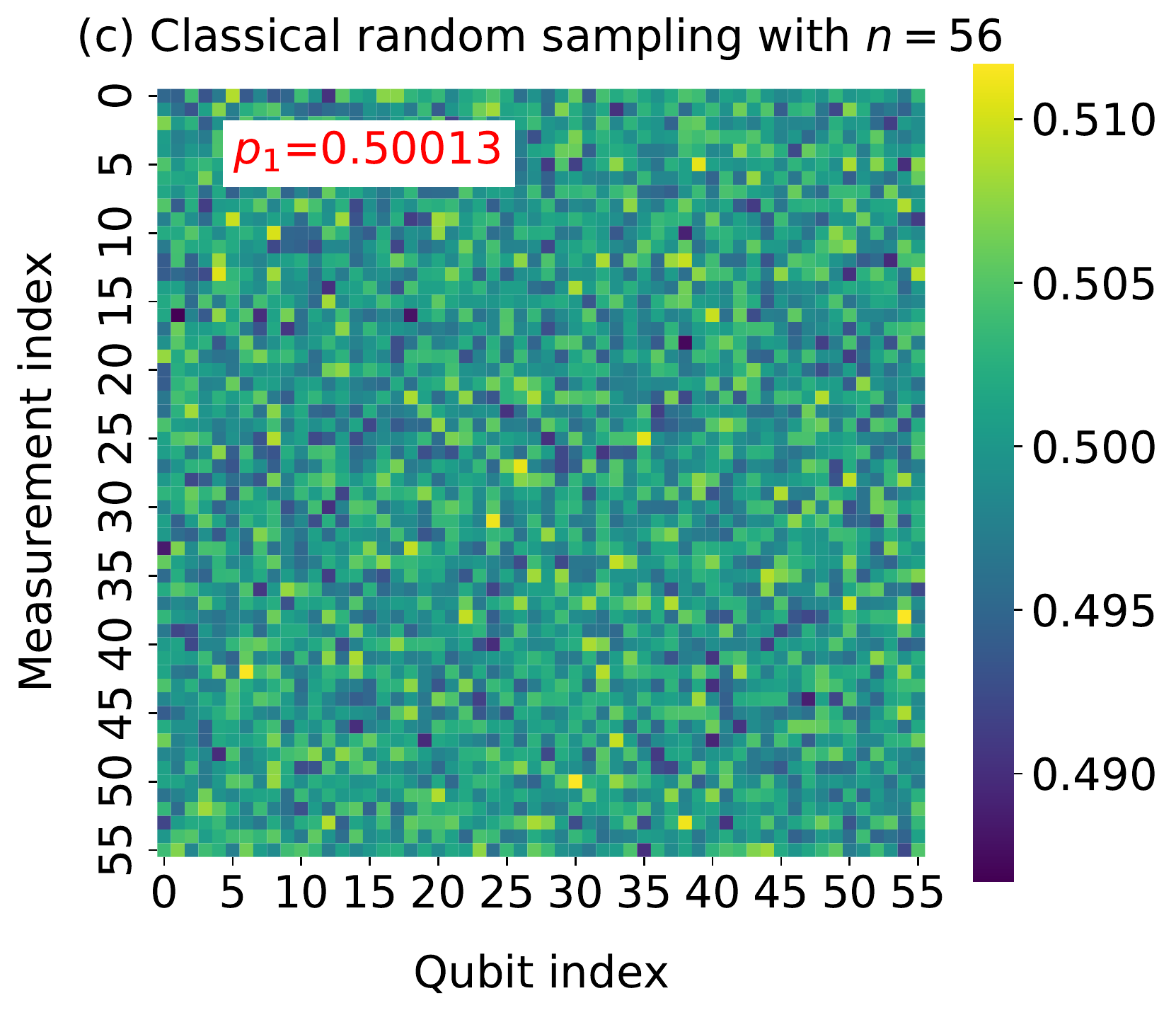}
\caption{\label{Fig1} Heatmaps of quantum random sampling (a) by  the Sycamore quantum processor with 
$n=53$ qubits, $m=20$ cycles, and sample numbers $M = 10^6$, (b) by the Zuchongzhi quantum processor 
with $n=56$ qubits, $m=20$ cycles, and $M = 10^6$ samples, and (c) of classical uniform random sampling 
with $n=56$ and $M=10^6$ are plotted. The average of finding bit 1 is denoted by $p_1$. Random quantum 
samples, (a) and (b), show the stripe patterns at specific qubit indices while (b) and (c) pass the NIST 
random number tests.}
\label{Fig1_heatmap}
\end{figure*}

To verify that a noisy quantum processor is performing well random quantum sampling~\cite{Arute2019,Wu2021}, 
the linear cross-entropy benchmarking (XEB) fidelity $F_{\rm XEB}$ is introduced
\begin{equation}
F_{\rm XEB} = 2^n\cdot\frac{1}{M}\sum_{x\in\cal D} p(x) -1\,,
\label{Eq:XEB}
\end{equation}
where the ideal probability $p(x) = |\bra{x}U\ket{0}|^2$ of finding a bit-string $x$ is computed on
a classical computer using Schr\"odinger or Feynman simulators and 
the bit-strings ${\cal D} = \{x_1,\dots,x_M\}$ are generated by a quantum processor. 
It is known that $F_{\rm XEB} =1$ if a quantum processor implements a random quantum circuit 
without errors, and $F_{\rm XEB} =0$ if bit-strings are sampled from a classical uniform 
distribution. The estimated XEB fidelity of Sycamore 
is $F_{\rm XEB}= (2.24\pm 0.021)\times 10^{-3}$ for 53 qubits, 20 
cycles, and $M = 30\times 10^6$ samples over 10 circuit instances~\cite{Arute2019}. 
For the Zuchongzhi quantum processor~\cite{Wu2021}, the estimated XEB fidelity is 
$F_{\rm XEB} = (6.62\pm 0.72)\times 10^{-4}$ for 56 qubits, 20 cycles, 
and $M=1.9\times 10^7$ samples.

While the XEB fidelity could serve as a benchmark for quantum advantage of random quantum sampling, 
it has some limitations. The XEB fidelity stems from the Kullback-Leibler divergence or the cross 
entropy of an empirical probability distribution $\tilde{p}(x)$ of the data ${\cal D}$ from the 
ideal probability distribution $p(x)$~\cite{Boixo2018}. As the number of qubits increases, it is very difficult 
to calculate both probability distributions. The number of samples required to construct 
the empirical probability $\tilde{p}(x)$ increases exponentially because the range of $x \in [0,2^n-1]$ 
does so.  While the XEB fidelity needs only an ideal probability $p(x)$ but not $\tilde{p}(x)$, 
it is still not scalable because in quantum advantage regime a supercomputer cannot calculate 
the ideal probability $p(x)$ in Eq.~{(\ref{Eq:XEB})} and no other NISQ processors could do. 
Recently, the limitation of the XEB fidelity as a benchmark for quantum advantage have been pointed 
out by Gao {\it et. al.}~\cite{Gao2021}. More importantly, the XEB fidelity could not give any 
clue about statistical properties of bit-strings of random quantum sampling implemented 
on NISQ processors.

\paragraph*{Comparison of random quantum sampling of Sycamore and Zuchongzhi.---}
A simple way of comparing the performance of two quantum processors is to compare directly their outputs
of a task, here random quantum sampling. Both Sycamore and Zuchongzhi have similar noise levels.  
The averages of single-qubit gate errors, two-qubit gates errors, and the readout error of Sycamore is 
about 0.15\%, 0.36\%, and 3.1\%, respectively~\cite{Arute2019}.  The average errors of Zuchongzhi are 
about 0.14\% for single-qubit gates, 0.59\% for two-qubit gates, and 4.52\% for readout, 
respectively~\cite{Wu2021}. Both processors implemented the same random quantum 
circuit as described before and obtained similar values of XEB fidelity. 
So, one may expect that their outputs would be statistically close each other.
However, that is not the case. 

\begin{figure*}[t]
\includegraphics[width=0.32\textwidth]{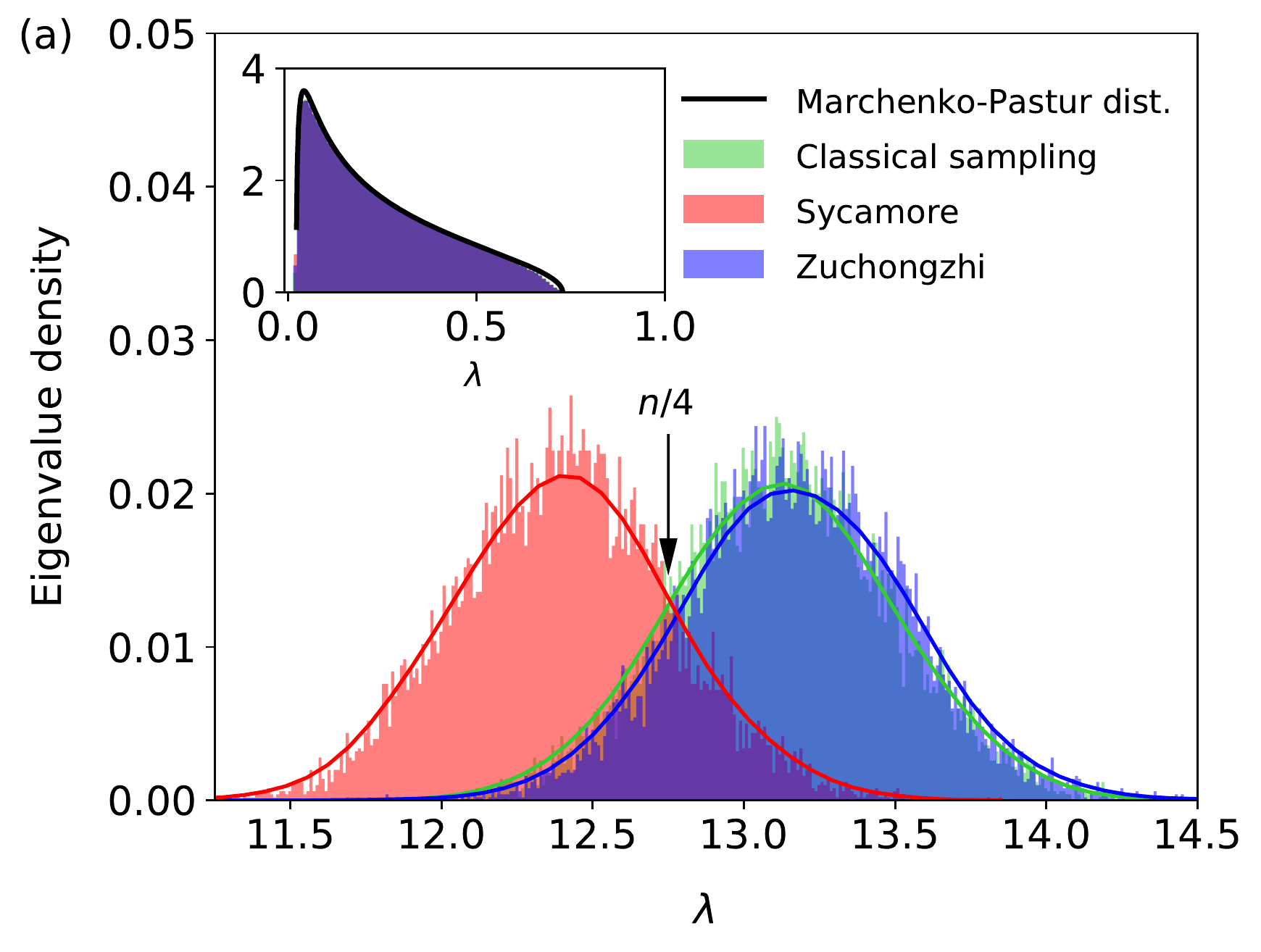}\hfill
\includegraphics[width=0.31\textwidth]{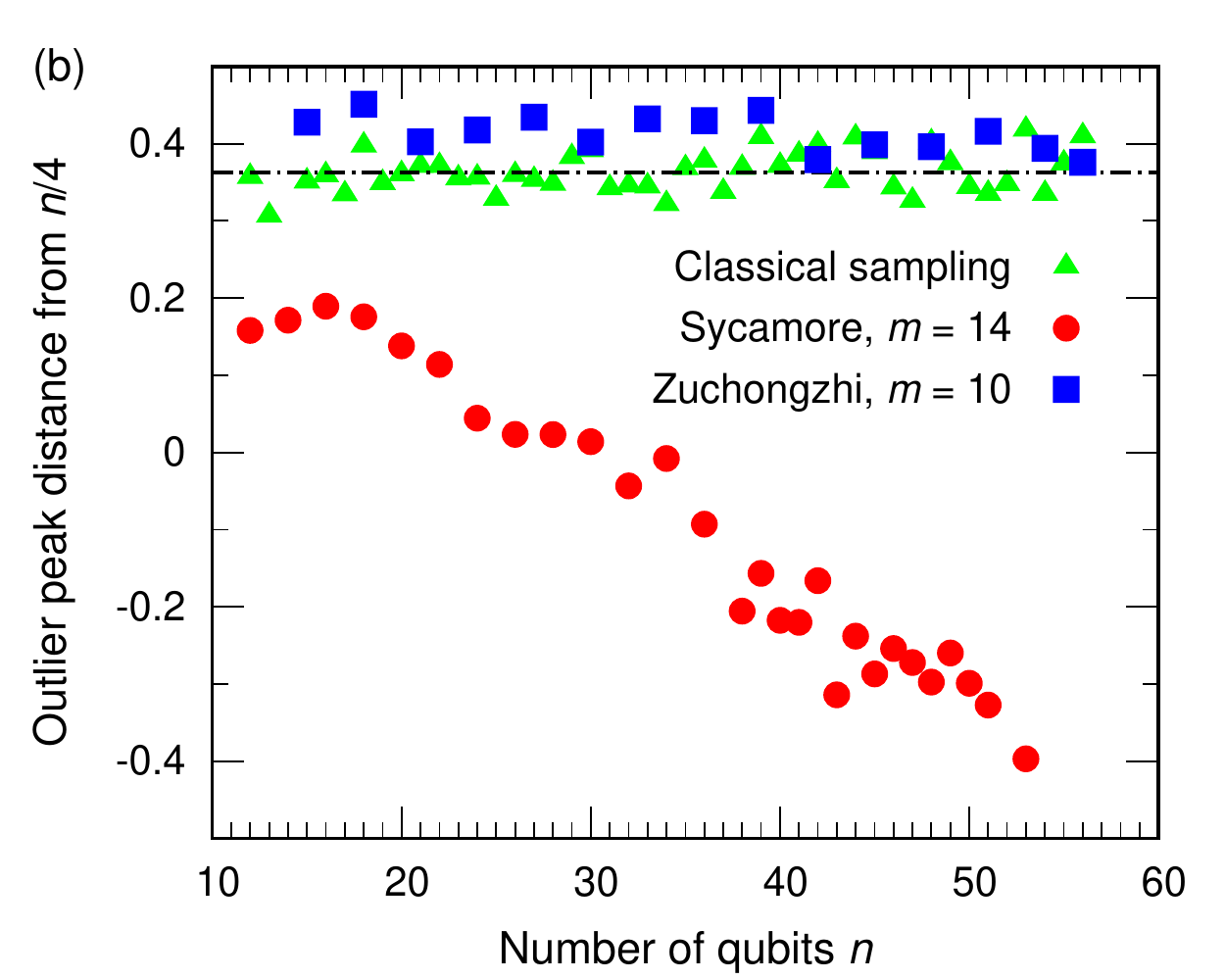}\hfill
\includegraphics[width=0.31\textwidth]{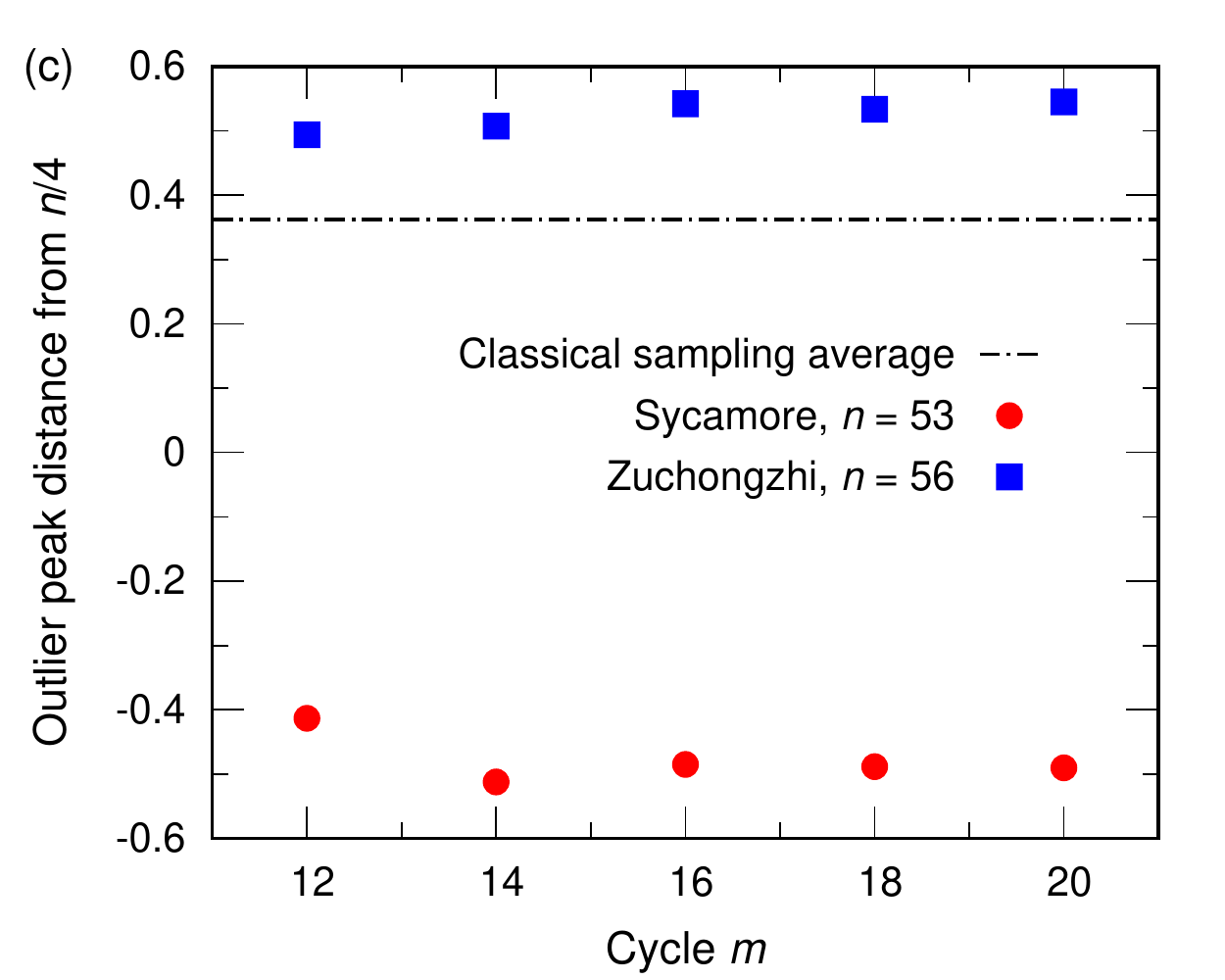}
\caption{(a) For $n=51$ qubits, the outliers of the Marchenko-Pastur distributions 
of eigenvalues of classical random bit-strings (green), Sycamore's quantum random bit-strings (red), 
and Zuchongzhi's quantum random bit-strings (blue) are plotted. 
The inset of (a) depicts the bulk distribution of eigenvalues of the Marchenko-Pastur law given 
by Eq.~(\ref{MP_eq}). (b) The distances of the outlier peaks of Sycamore (red solid circle ), 
Zuchongzhi (blue solid square), and  classical random sampling (green triangle) from $n/4$ are plotted 
as a function of the qubit number $n$. (c) The distances of the peaks of Sycamore with $n=53$ 
(solid red circle) and Zuchongzhi with $n=56$ (blue solid square) from $n/4$ is plotted as 
a function of cycle $m$.  The dot-dash lines in (b) and (c) represent the average distance
of classical random sampling from $n/4$. 
}
\label{Fig2_MP}
\end{figure*}

We first examine whether the output bit-strings of random quantum sampling are random. 
Random numbers have many practical applications in Monte-Carlo simulation, statistics, cryptography, 
etc.. Given a random quantum circuit $U$, the probability $p(x)=|\bra{x}U\ket{0}|^2$ of finding bit-string $x$ 
is not uniform due to the entanglement and interference, so some bit-strings are sampled more likely 
than others. However, there is no physical ground that the output of bit-strings ${\cal D}=\{x_1,x_2,\dots,x_M\}$
would contain more bit 1 than 0 or vice versa. Bits $0$ and $1$ are equally probable if there is no error. 
To see this, the output of bit-strings ${\cal D}$ is sliced into the collection of $n\times n$ binary arrays, 
with which the heatmaps are plotted as shown in Fig.~\ref{Fig1_heatmap}.
Fig.~\ref{Fig1_heatmap}~{(a)} depicts the heat map of Sycamore's bit-strings for $n=53$ qubits, $m=20$ cycles,
and $M=3\times10^6$ samples~\cite{Martinis2021}. 
Fig.~\ref{Fig1_heatmap}~{(b)} displays the heatmap of Zuchongzhi's bit-strings for $n=56$ 
qubits, $m=18$ cycles, and $M=3\times 10^6$ samples. Fig.~\ref{Fig1_heatmap}~{(c)} shows the heatmap of uniform 
random bits sampled from a classical computer for $n=56$ and $M=3\times10^6$. As depicted in Fig.~\ref{Fig1}, 
the heatmaps of random quantum sampling on Sycamore and Zuchongzhi quantum processors show bright and darks 
stripes at some qubit indices while the classical uniform random sampling does not. One may suspect that stripe 
patterns could be caused by readout errors. However, for both Sycamore and Zuchongzhi data sets, 
the locations of bright or dark stripes in the heatmaps do not coincide with the indices of qubits with 
high readout errors.

We count the number of bit $1$ contained in the output bit-strings ${\cal D} = \{x_1,\cdots,x_M\}$.
The average of finding bit 1 is denoted by $p_1$. Most Zuchongzhi outputs show $p_1$ is greater 
than 1/2, as shown in Supplementary Material~\cite{Suppl2022}, while all Sycamore outputs have $p_1$ 
less than 1/2~\cite{Oh2021}. As depicted in Fig.~\ref{Fig1}, Zuchongzhi's bit-strings with $n=56$ 
qubits, $m=18$ cycles has $p_1 = 0.50094$ while Sycamore's bit-strings for $n=53$ qubits and $m=20$ 
cycles has $p_1 = 0.48383$. Note that the readout errors of Sycamore and Zuchongzhi are reported 
to be 3.1\% and 4.52\%, respectively. It is unclear why Sycamore and Zuchongzhi 
are quite different in the probability $p_1$ of finding bit $1$ in random quantum sampling.

To see the randomness of bit-strings, we perform the NIST random number tests~\cite{NIST2010} 
for Zuchongzhi's outputs. We find that some Zuchongzhi data sets pass the NIST random number tests 
while all Sycamore data sets do not~\cite{Suppl2022,Oh2021}. Note that the probability $p_1$ of 
finding bit 1 is related to the frequency test of the NIST random number tests.
It is interesting to see that some Zuchongzhi data sets pass the NIST random number tests while 
all random quantum sampling data of Sycamore and Zuchongzhi show the stripe patterns unlike 
the classical uniform random samples, as shown in Fig.~\ref{Fig1}. This implies random quantum 
sampling may be used to generate quantum random numbers~\cite{Collantes2017,Tamura2021}. 


Both Sycamore and Zuchongzhi showed similar exponential decays of the XEB fidelity $F_{\rm XEB}$ 
as a function of qubit number $n$ or cycle $m$~\cite{Arute2019,Wu2021}. One obtains $F_{\rm XEB} =0$
if bit-strings are sampled from the classical uniform random distribution $p_{\rm cl}(x) = 1/2^n$.  
Using the Marchenko-Pastur distribution~\cite{MarPas67} and the Wasserstein 
distances~\cite{Villani2008,flamary2021pot}, we examine how the random bit-string outputs of Sycamore 
and Zuchongzhi are far away from classical uniform random samples as a function of qubit number $n$ 
and cycle $m$. 


The signal out of randomness can be captured by the outliers of the Marchenko-Pastur distribution of 
random matrices. This method has been applied to the covariance matrices in finance~\cite{Laloux1999,Laloux2000} 
in predicting ligand affinity~\cite{Lee2016}, and in denoising MRI~\cite{Veraart2016} and single-cell 
data~\cite{Aparicio2020}.
We use the Marchenko-Pastur distribution of eigenvalues of random bit-strings to measure the distance
among Sycamore's outputs, Zuchongzhi's output, and classical uniform random samples.
The data ${\cal D} = \{x_1,x_2,\dots,x_M\}$, an binary $M\times n$ array, is sliced into the 
collection of $k\times n$ random matrices $X$. All entries of $X$ are assumed to be independent 
and identically binary random variable $\{0,1\}$ with the mean $\mu_X = 1/2$ and
the variance $\sigma_X^2 = 1/4$. We calculate the empirical distribution of eigenvalues of 
$n\times n$ matrix $\frac{1}{k}X^t\cdot X$, as shown in Fig.~\ref{Fig2_MP}.
The empirical eigenvalue distribution is composed of the two parts: the bulk 
corresponding to the noise or random and the outliers representing the signal. 
To understand this, let consider the transformed random matrix $Y = 2X - J$ with $J$ 
of all entries 1.  The matrix $\frac{1}{k}X^t\cdot X$ is written as
\begin{align}
\frac{1}{k}X^t\cdot X = \frac{1}{4k}\left(Y^t\cdot Y + X^t\cdot J + X^t\cdot Z + J^t\cdot J\right).
\label{Wishart}
\end{align}
The matrix $Y$ has the mean $\mu_Y =0$ and the variance $\sigma^2_Y = 1$. The eigenvalue distribution of 
the first term in Eq.~(\ref{Wishart}), $\frac{1}{k} Y^t\cdot Y$, follows the Marchenko-Pastur 
distribution~\cite{MarPas67}
\begin{align}
\rho(\lambda) = \frac{1}{2\pi\sigma^2 \gamma} \frac{\sqrt{(\lambda_+ -\lambda)(\lambda -\lambda_{-})}}{\lambda}\,,
\label{MP_eq}
\end{align}
where $\gamma = n/k$ is the rectangular ratio and $\lambda_{\pm} = \sigma^2(1 \pm\sqrt{\gamma})^2$ are the upper 
and lower bounds. Here we take $k=2n$, i.e., $\gamma = 1/2$. The upper limit is given by 
$\lambda_+ = 1 +\sqrt{1/2}$. By considering the scaling factor $1/4$, the upper limit 
of $(1/4k)\cdot Y^t\cdot Y$ is given by $(1 +\sqrt{1/2})/4\approx 0.7285$. 
The last term of Eq.~(\ref{Wishart}), $(1/4k)J^t\cdot J$ has the two eigenvalues, 0 and $n/4$. 
So the outliers are located around $n/4$.

Fig.~\ref{Fig2_MP} (a) plots the empirical distribution of eigenvalues of $\frac{1}{k}X^t\cdot X$ made of
Sycamore bit-strings with $n=51$, $m =14$, and $M=10^6$ (red color),  Zuchongzhi bit-strings with $n=51$, 
$m=10$ and $M=10^6$ (blue color), and classical uniform random bits with $M=10^6$ (green color). 
The outlier peak of Sycamore is located at the left of $n/4$ while the outlier peaks of Zuchongzhi and 
classical random sampling are located at right of $n/4$. Surprisingly, Zuchongzhi's bit-strings are more closer 
to the classical random bit strings than Sycamore's ones while Sycamore and Zuchongzhi have similar the values
of the XEB fidelity. Fig.~\ref{Fig2_MP} (b) plots the distances of the outlier peaks from $n/4$ as a function
of qubit number $n$. Sycamore becomes farther away from $n/4$ as $n$ increases. On the other hands,
Zuchongzhi's distance from $n/4$ changes a little. The XEB fidelity of Sycamore and Zuchongzhi decays
exponentially as a function of qubit number $n$. Fig.~\ref{Fig2_MP} (c) plots the peak distance from $n/4$
as a function of cycles for Sycamore $n=53$ and for Zuchongzhi $n=56$. This result is also in contrast 
with the decay behavior of the XEB fidelity as a function of cycle $m$.

\begin{figure}[t]
\includegraphics[width=0.45\textwidth]{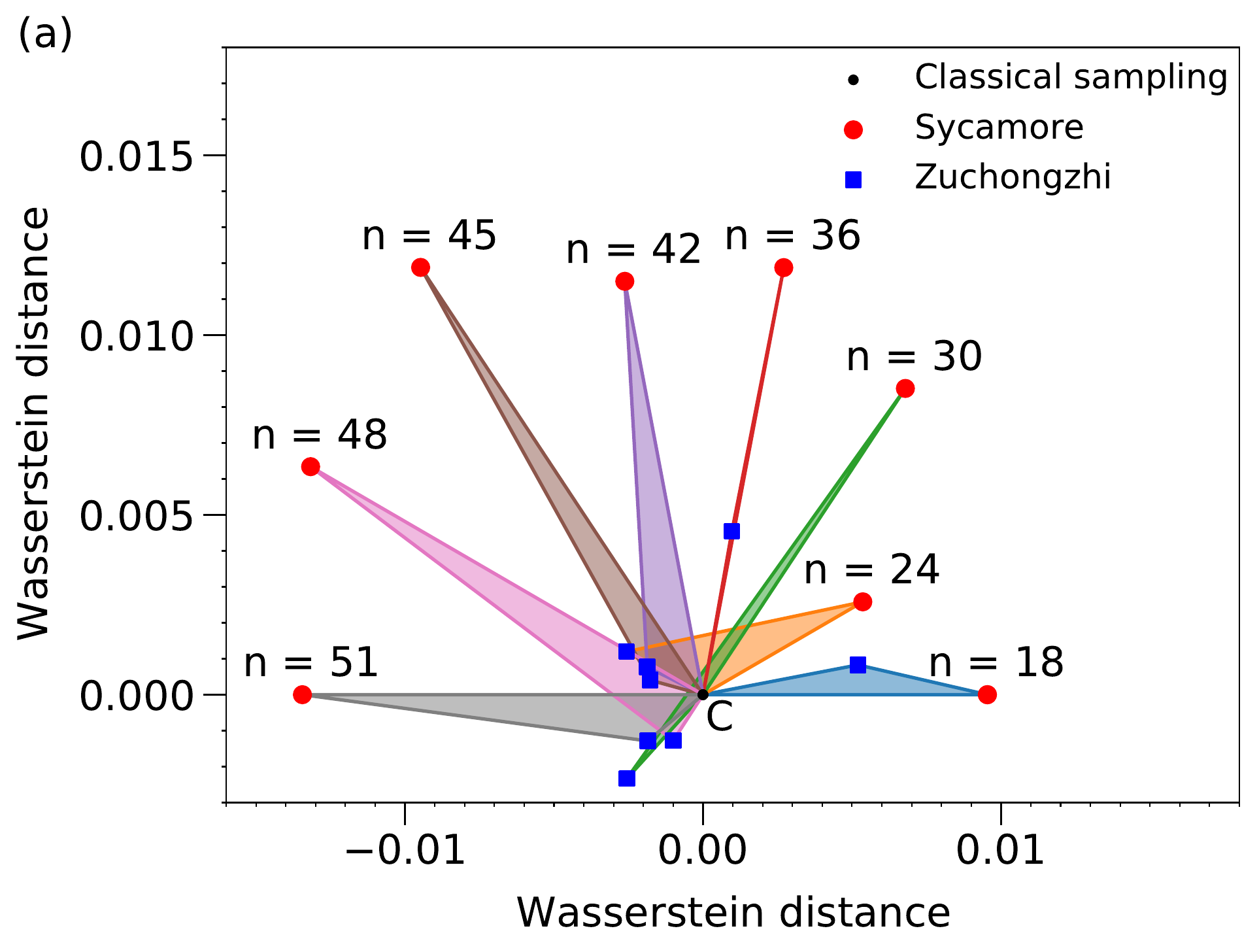}
\includegraphics[width=0.45\textwidth]{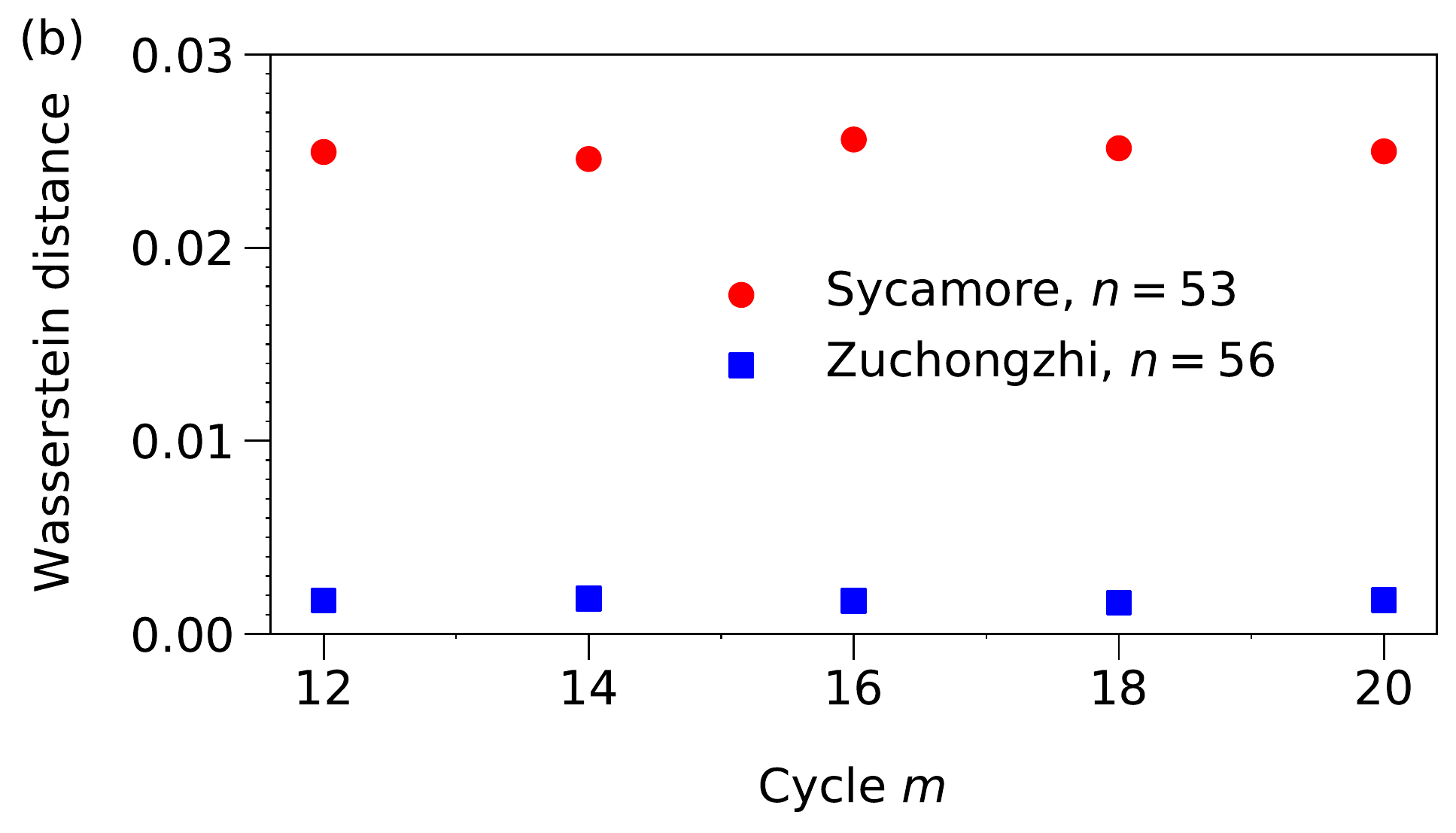}
\caption{(a) For qubits $n=18,24, 30, 36, 42, 45, 48, 51$, the relative positions of Sycamore (red circle), 
Zuchongzhi (blue square), and classical random sampling (black circle at origin) form triangles whose 
edges are the Wasserstein distances. For clarity, triangles are rotated.
(b) for Sycamore bit-strings with $n=51$ and Zuchongzhi with $n=56$, the Wasserstein distances 
from the classical random sampling are plotted as a function of cycle $m$.
}
\label{Fig_WD1}
\end{figure}

The statistical distances among Sycamore, Zuchongzhi, and classical random sampling, based on 
the Marchenko-Pastur distribution, can be further confirmed by calculating the Wasserstein distances.
The 1-th Wasserstein distance between two probability distribution $p(x)$ and $q(x)$~\cite{Villani2008}
is defined by 
\begin{align}
W_p(p,q) =  \inf_{\pi\in \Pi(p,q)} {\mathbb E}_{(x,y)\sim \pi} \left[||x - y ||\right] \,,
\end{align}
where $\Pi(p,q)$ is the set of all joint distribution $\pi(x,y)$ whose
marginal distributions  are $p(x)$ and $p(y)$, respectively.
Given two samples, $\{x_1,x_2,\dots,x_M\}$ and
$\{y_1,y_2,\dots,y_M\}$, $W(p,q)$ can be calculated directly without calculating the empirical distributions
$p(x)$ and $q(x)$~\cite{flamary2021pot}. The Wasserstein distance is a true metric on a probability space,
so the relative Wasserstein distances among Sycamore, Zuchongzhi, 
and classical random sampling, give rise to the triangle inequality. Fig.~\ref{Fig_WD1} (a) plot the triangle 
relation among the three data sets for $n=18,24,30,36,42,45,48,51$. All Zuchongzhi data sets are close to
classical random sampling than Sycamore. This is consistent with the outlier peak distances from $n/4$ of 
the Marchenko-Pastur distribution shown in Fig.~\ref{Fig2_MP} (b). Fig.~\ref{Fig_WD1} (b) plots the Wasserstein 
distance of Sycamore $n=53$ and Zuchongzhi $n=56$ from the classical random bit-strings as a function of cycle 
$m$. This result is consistent with the behavior of the relative distances of the outliers of the Marchenko-Paster 
distribution as a function of $m$, shown in Fig.~\ref{Fig2_MP} (c). Note that the Wasserstein distances between 
the bit-strings with different cycles of Sycamore (or Zuchongzhi) are less than the Wasserstein distance 
of them from the classical random sampling.

\paragraph*{Conclusion.---}
In this paper, we analyzed the statistical properties of random bit-strings sampled from Sycamore and Zuchongzhi 
quantum processors using the heatmaps, the NIST random number tests, the Marchenko-Pastur distribution, and 
the Wasserstein distances. Both Sycamore and Zuchongzhi exhibit stripe patterns in the heatmaps of random 
bit-strings. Some of Zuchongzhi's data pass the NIST random number tests. This may open up a possibility
of the use of the random quantum circuit as a quantum random number generator. The Marchenko-Pastur 
distribution and the Wasserstein distances of random bit-strings shows that Zuchongzhi random quantum sampling 
is statistically more closer to classical random sampling than Sycamore's, while both have similar error rates and
similar values of the XEB fidelity. The distances of random bit-strings of Sycamore from the classical uniform 
sampling as a function of the number of qubit $n$ is quite different from those of Zuchongzhi.
The distances of Sycamore's and Zuchongzhi's bit-strings from the classical random bit-strings remain
almost constant as the cycle number $m$ increases. This is in contrast with the exponential decay of 
the XEB fidelity as a function of $m$.

The results found here raise a question about computational reliability and verification of noisy quantum 
processors~\cite{Eisert2020}. Two classical computers performing the same task will produce the same outputs 
(within statistical 
errors for stochastic calculation). One may expect that two NISQ processors with similar error rates 
would produce statistically close outputs. However, the results here show an exceptional case.
In order to ensure the reliability of quantum computing with NISQ quantum processors, it is necessary to
have good benchmark tools which can verify the output of quantum calculation on a classical computer,
until fully-error corrected quantum computers are available.

\begin{acknowledgments}
We would like to thank  the Google quantum team for making their random quantum sampling by Sycamore available.
We also would like to thank Y. Wu and J.-W. Pan for providing us the data on Zuchongzhi~\cite{Wu2021}.  
This material is based upon work supported by the U.S. Department of Energy, Office of Science, National 
Quantum Information Science Research Centers. We also acknowledge the National Science Foundation under 
award number 1955907. 
\end{acknowledgments}

\bibliography{main_oh_kais}
\clearpage




\begin{widetext}
\section*{Supplementary material: \\ Statistical Analysis on Random Quantum Sampling by Sycamore and Zuchongzhi 
Quantum Processors}
\subsection*{Sangchul Oh and Sabre Kais}
%


Supplemental material presents the heatmaps and the NIST random number tests for random bit-strings of Zuchongzhi.
The Wasserstein distances among different cycles for Sycamore and Zuchongzhi are presented.

\begin{figure*}[h]
\includegraphics[width=0.32\textwidth]{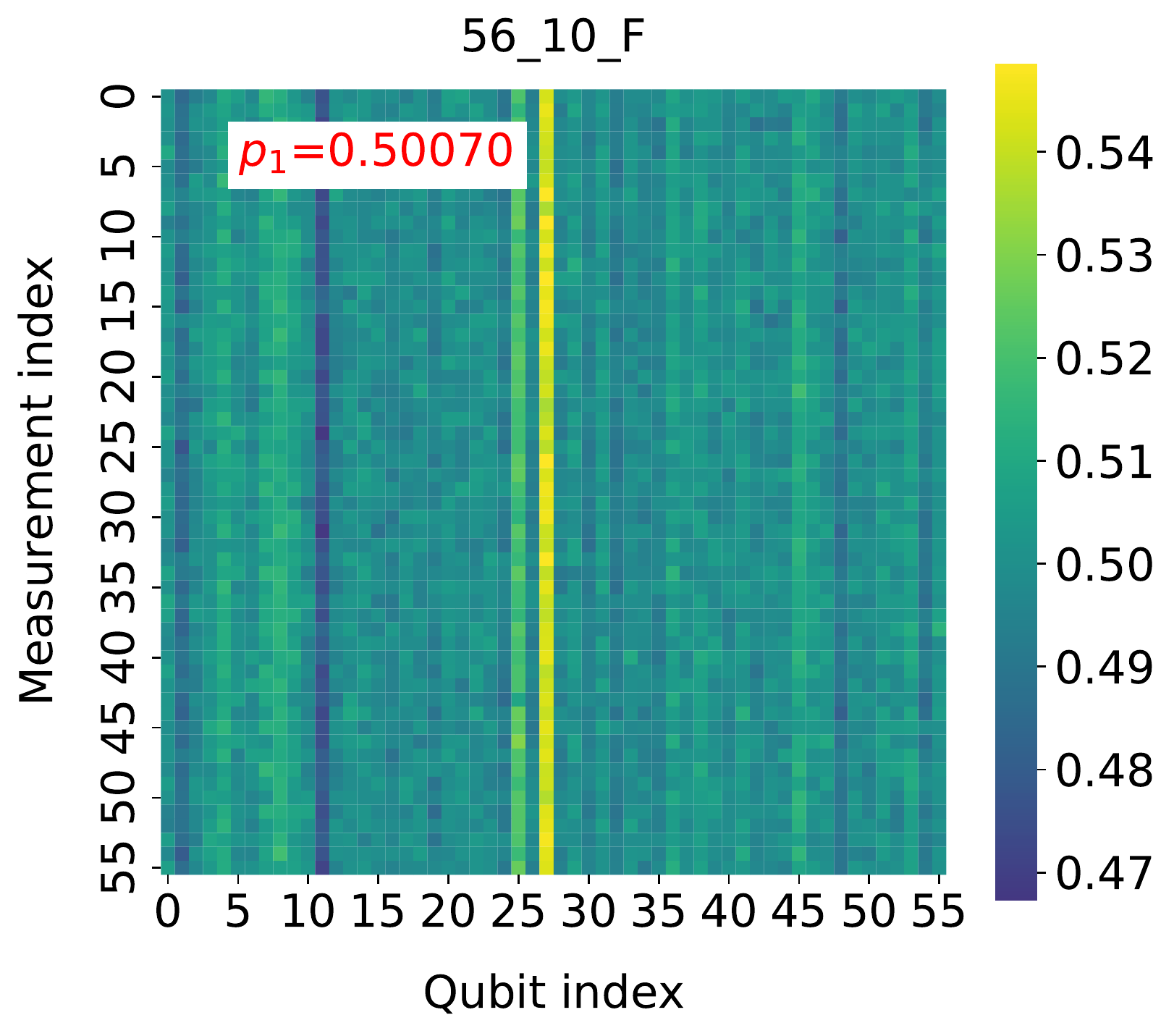}
\includegraphics[width=0.32\textwidth]{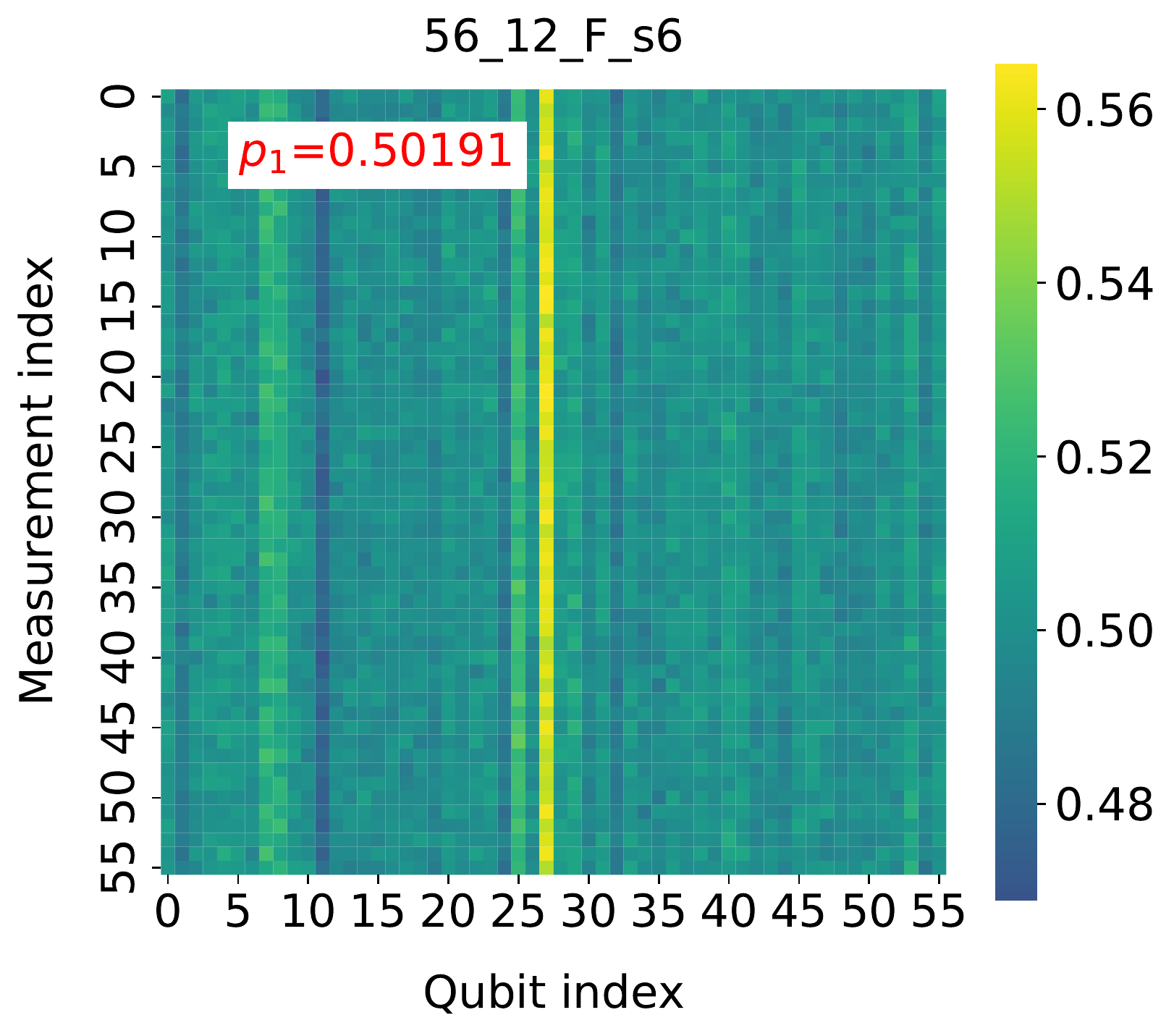}
\includegraphics[width=0.32\textwidth]{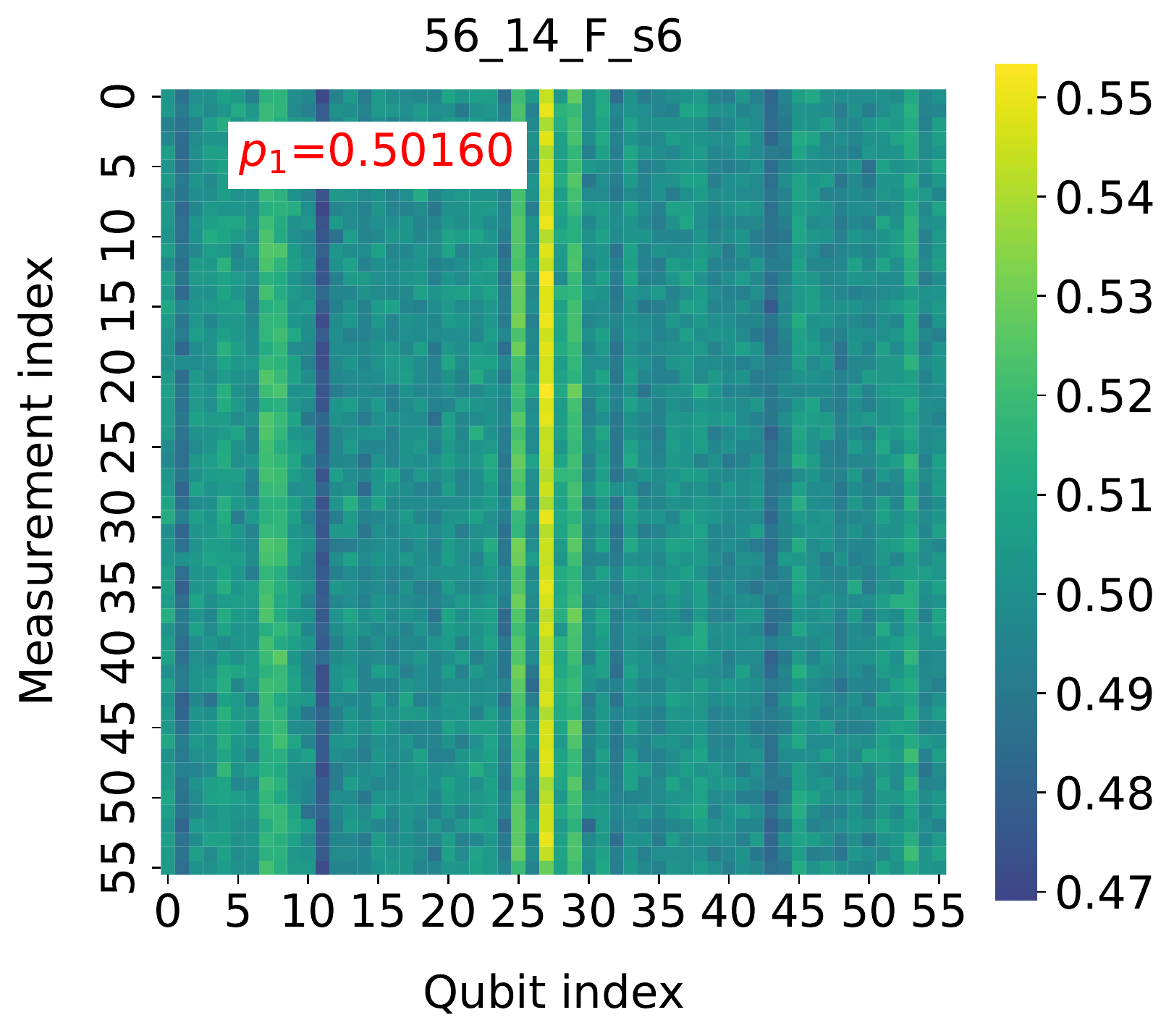}
\includegraphics[width=0.32\textwidth]{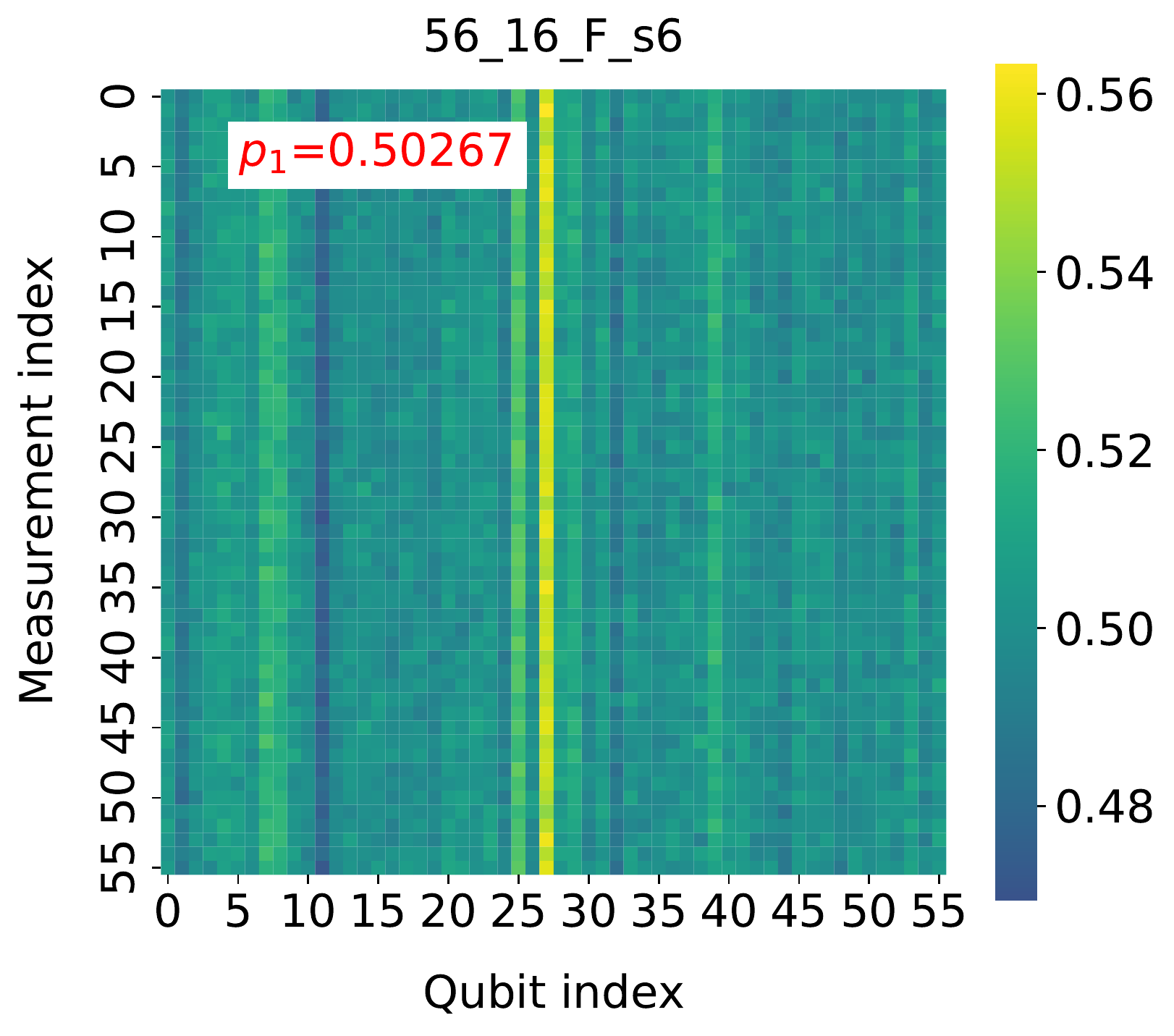}
\includegraphics[width=0.32\textwidth]{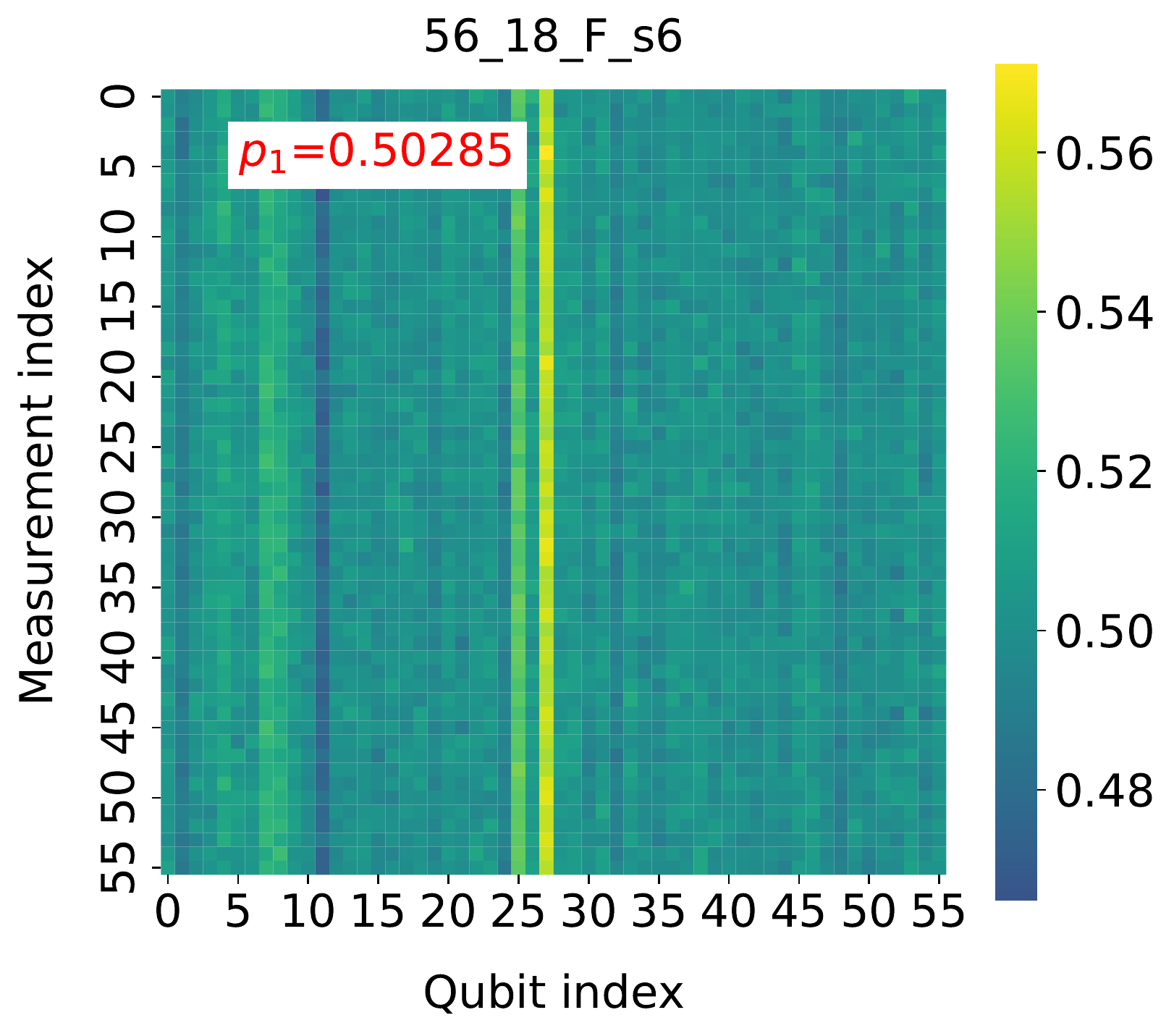}
\includegraphics[width=0.32\textwidth]{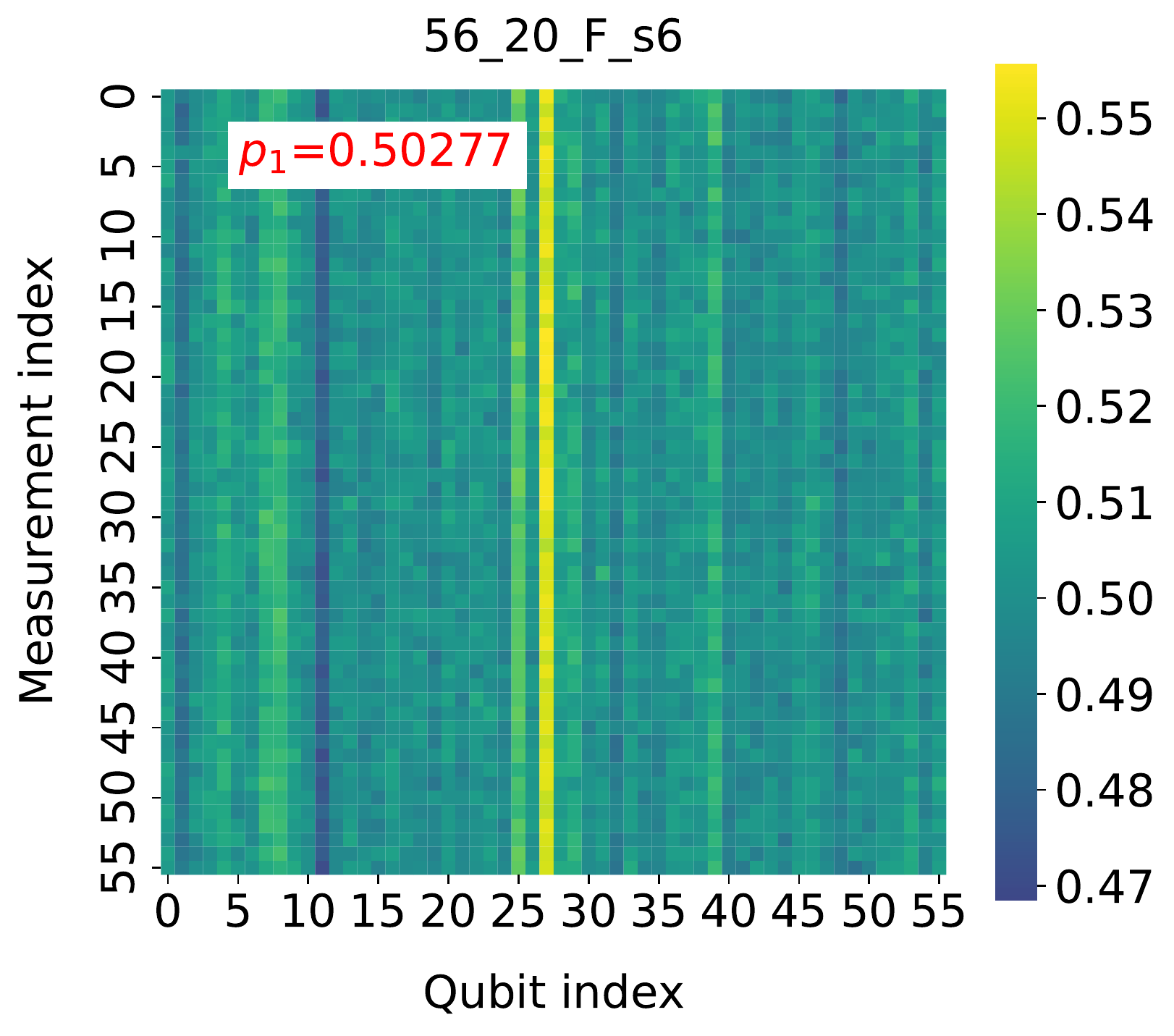}
\caption{Heatmaps of random bit-strings of Zuchongzhi with $n=56$ and $m=10,12,14,16,18,20$.}
\label{S_Fig1}
\end{figure*}

\begin{figure*}[t]
\includegraphics[width=0.25\textwidth]{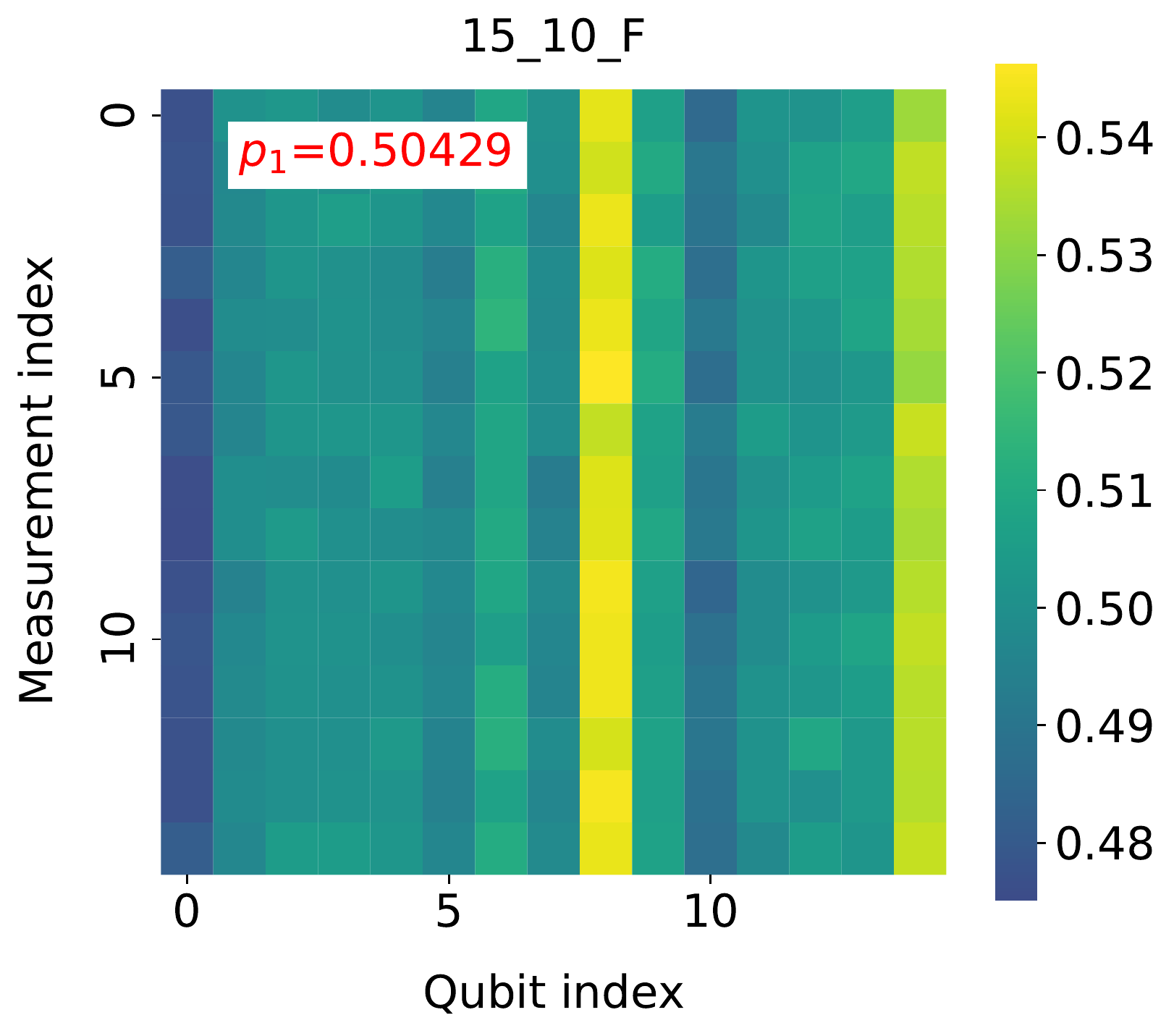}
\includegraphics[width=0.25\textwidth]{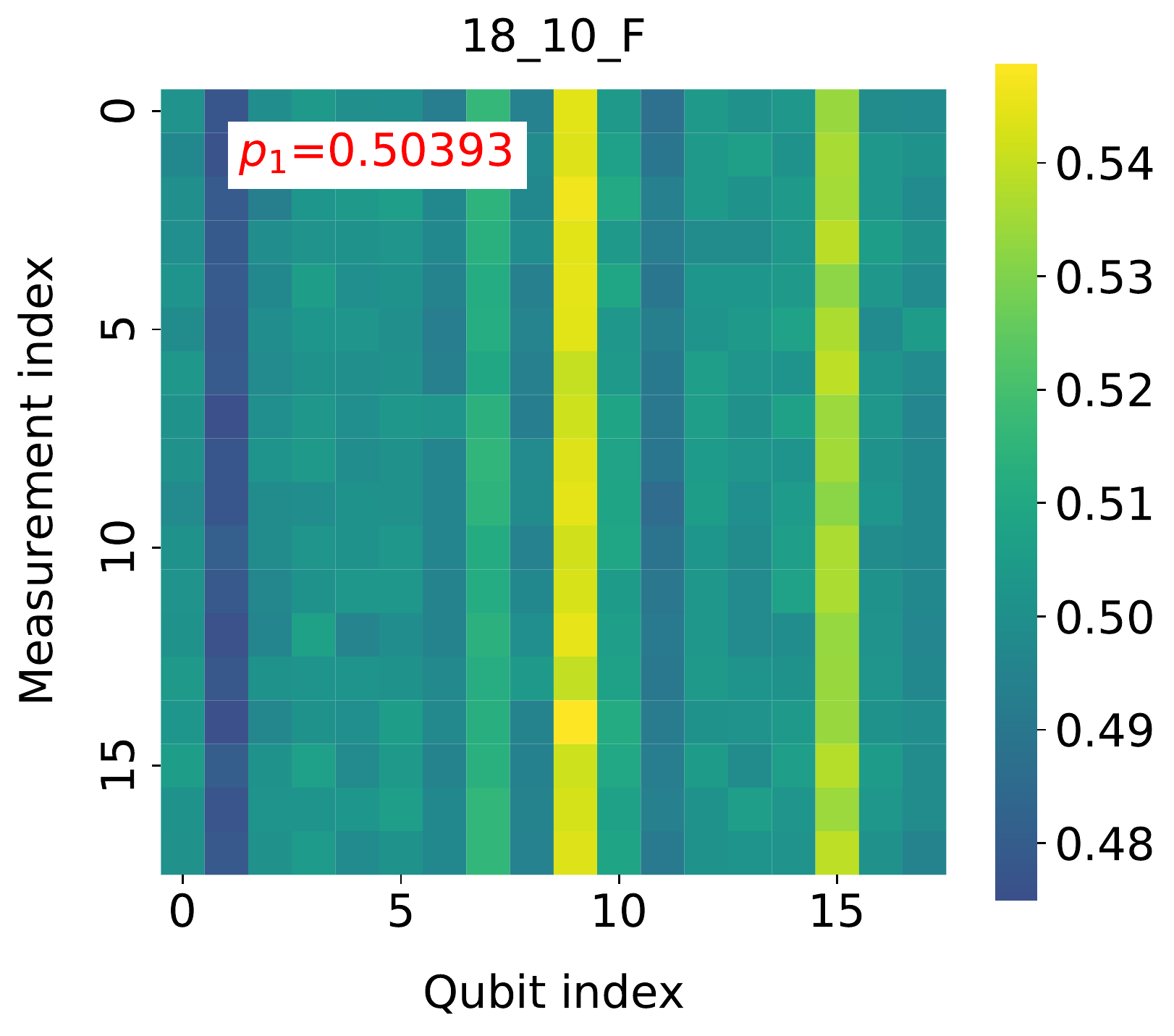}
\includegraphics[width=0.25\textwidth]{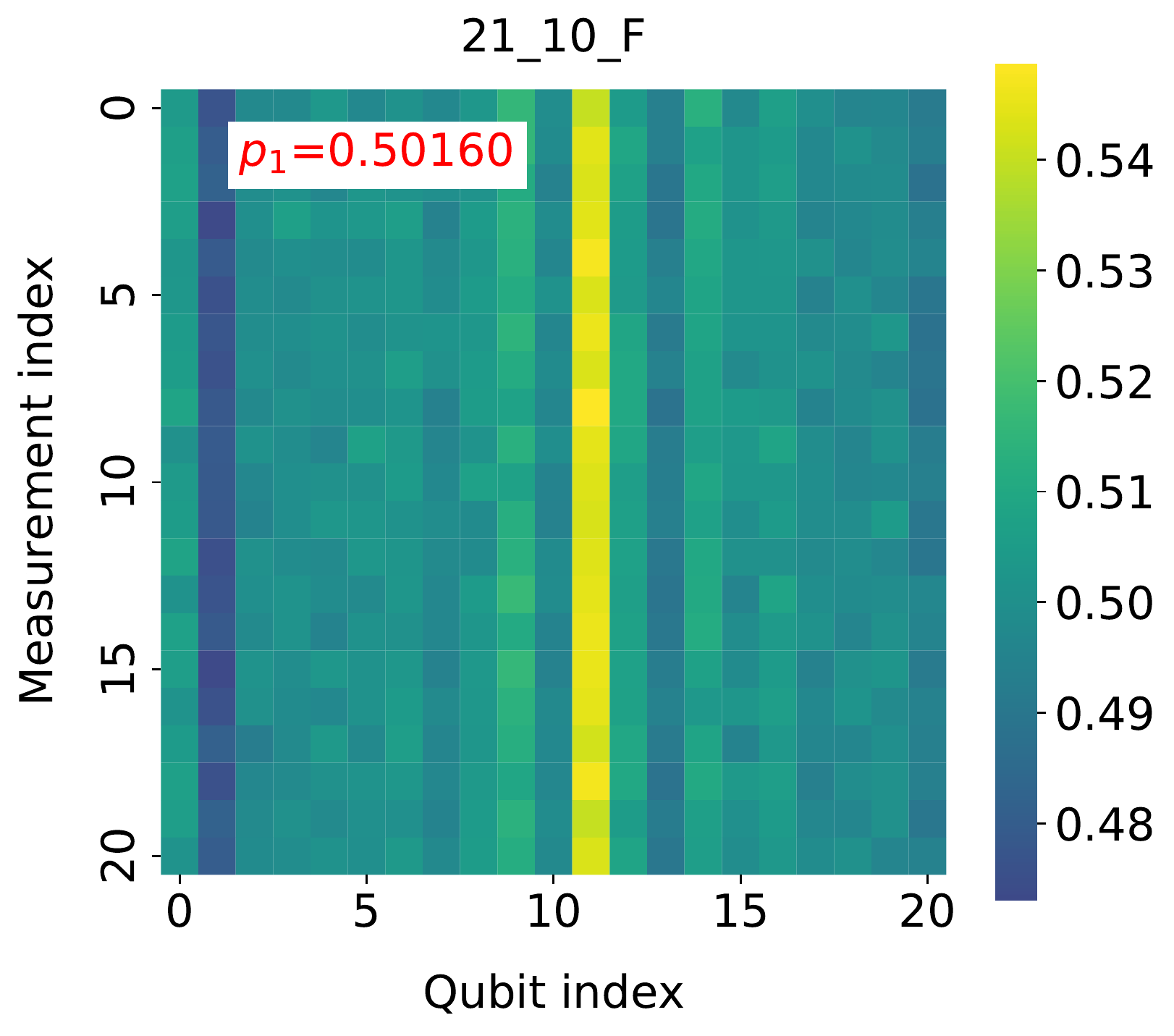}\\
\includegraphics[width=0.25\textwidth]{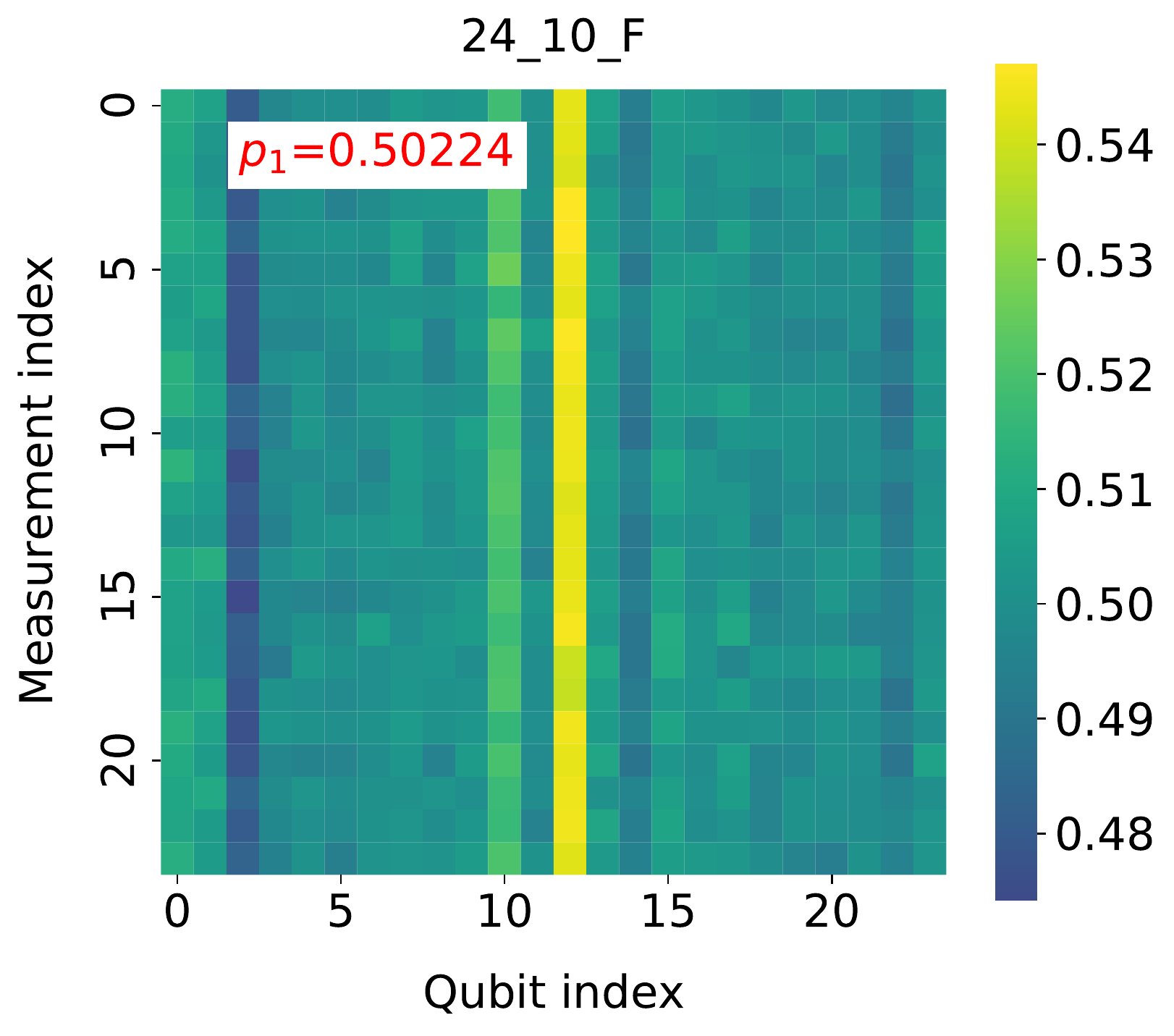}
\includegraphics[width=0.25\textwidth]{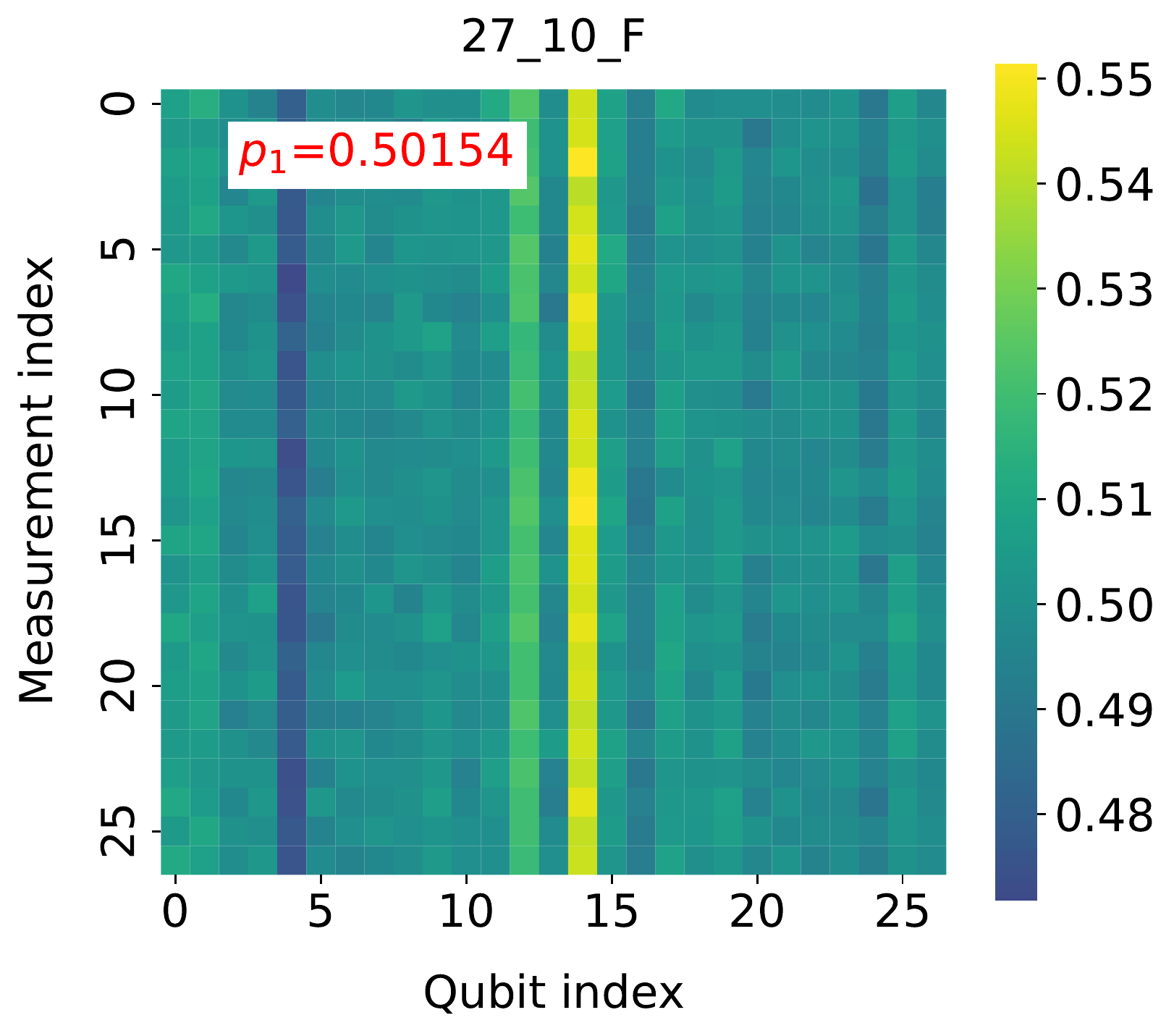}
\includegraphics[width=0.25\textwidth]{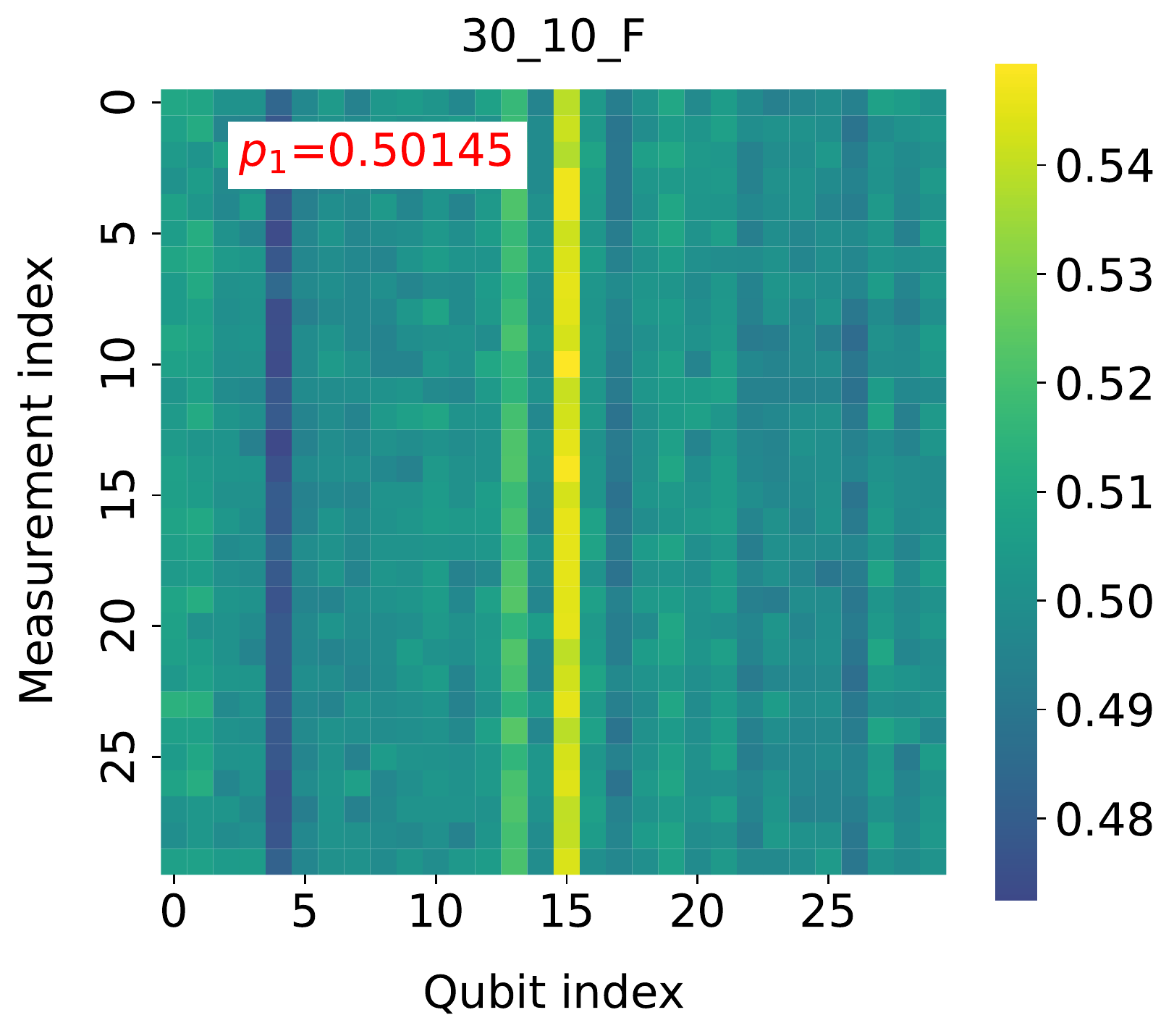}\\
\includegraphics[width=0.25\textwidth]{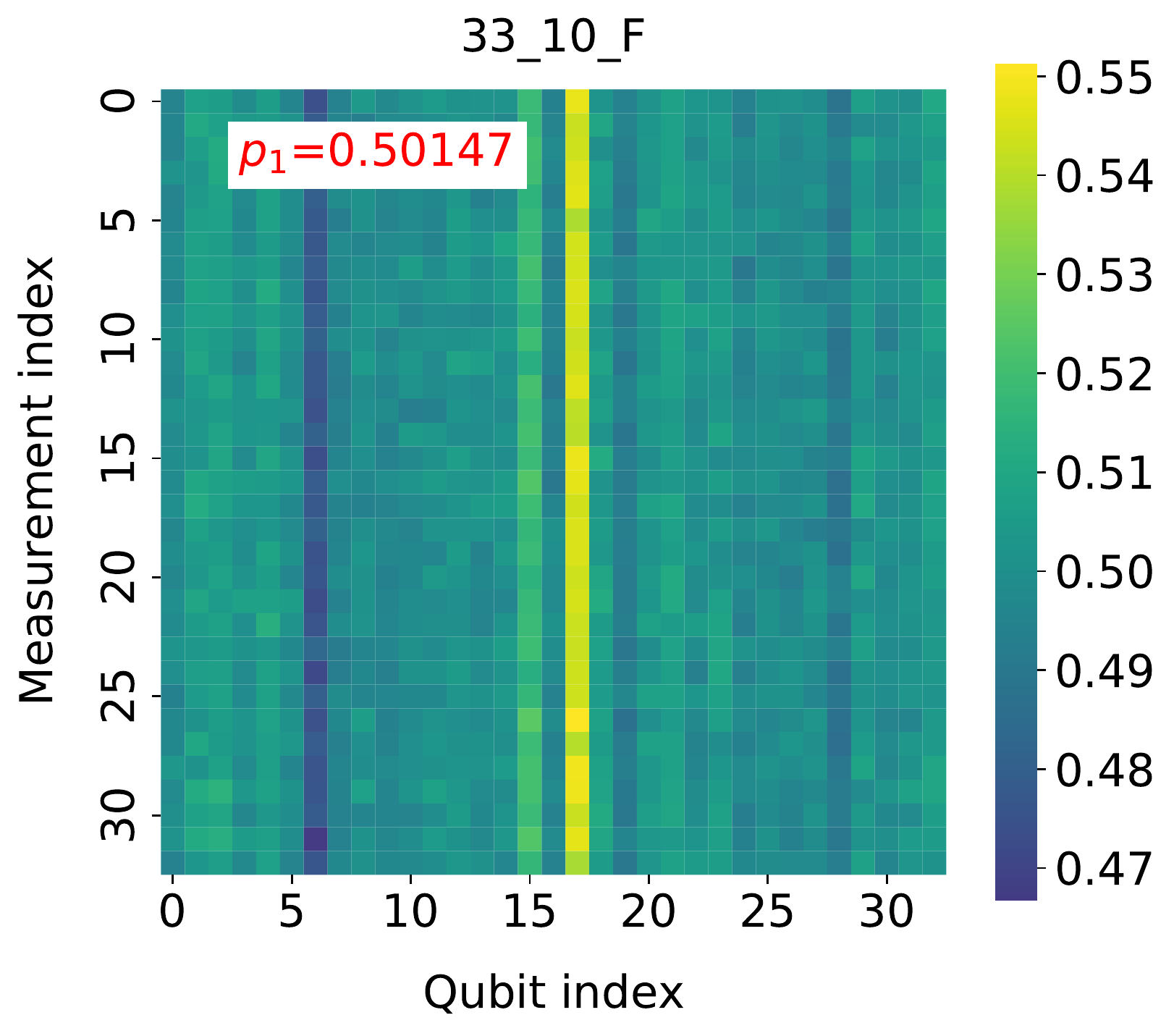}
\includegraphics[width=0.25\textwidth]{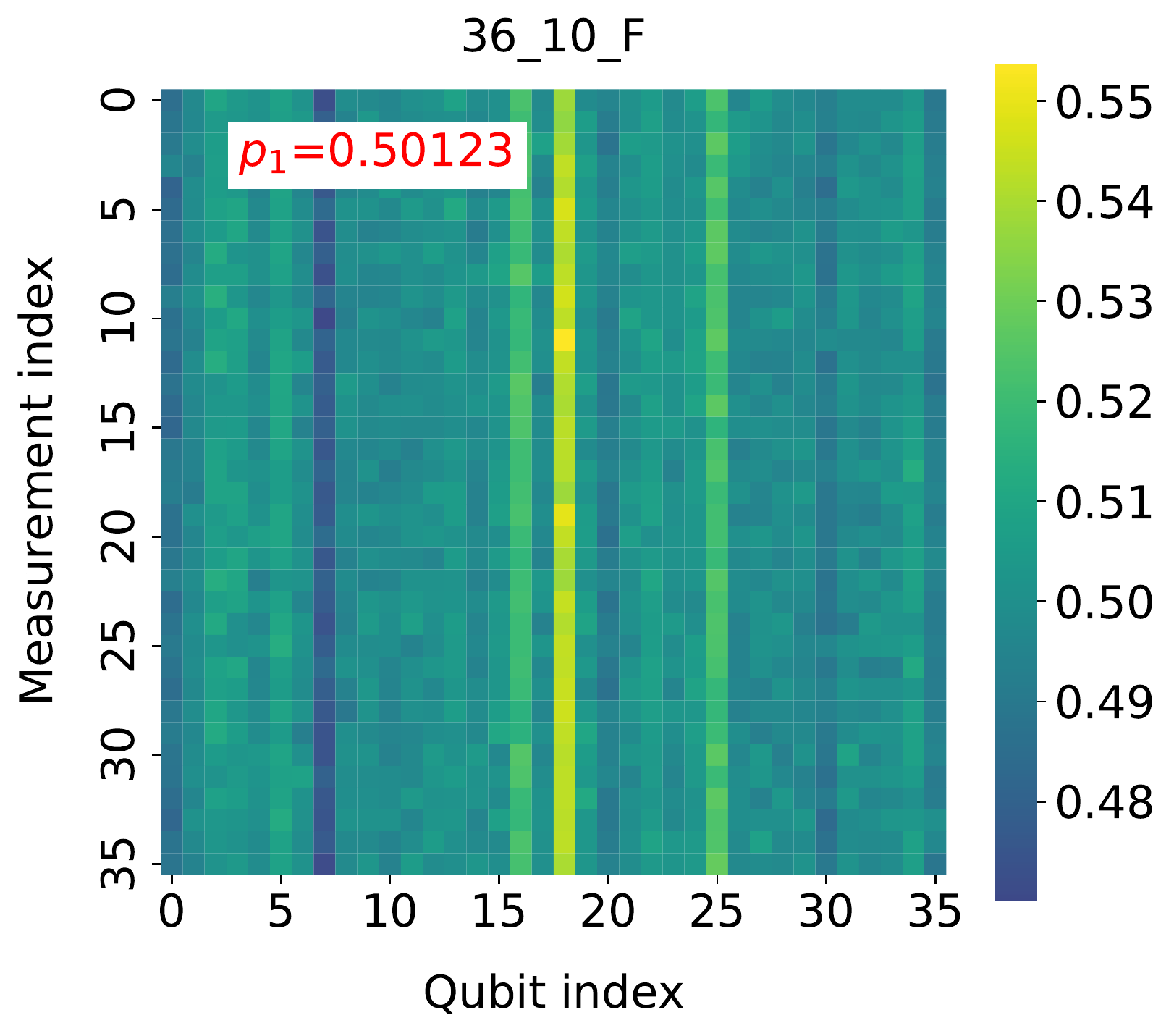}
\includegraphics[width=0.25\textwidth]{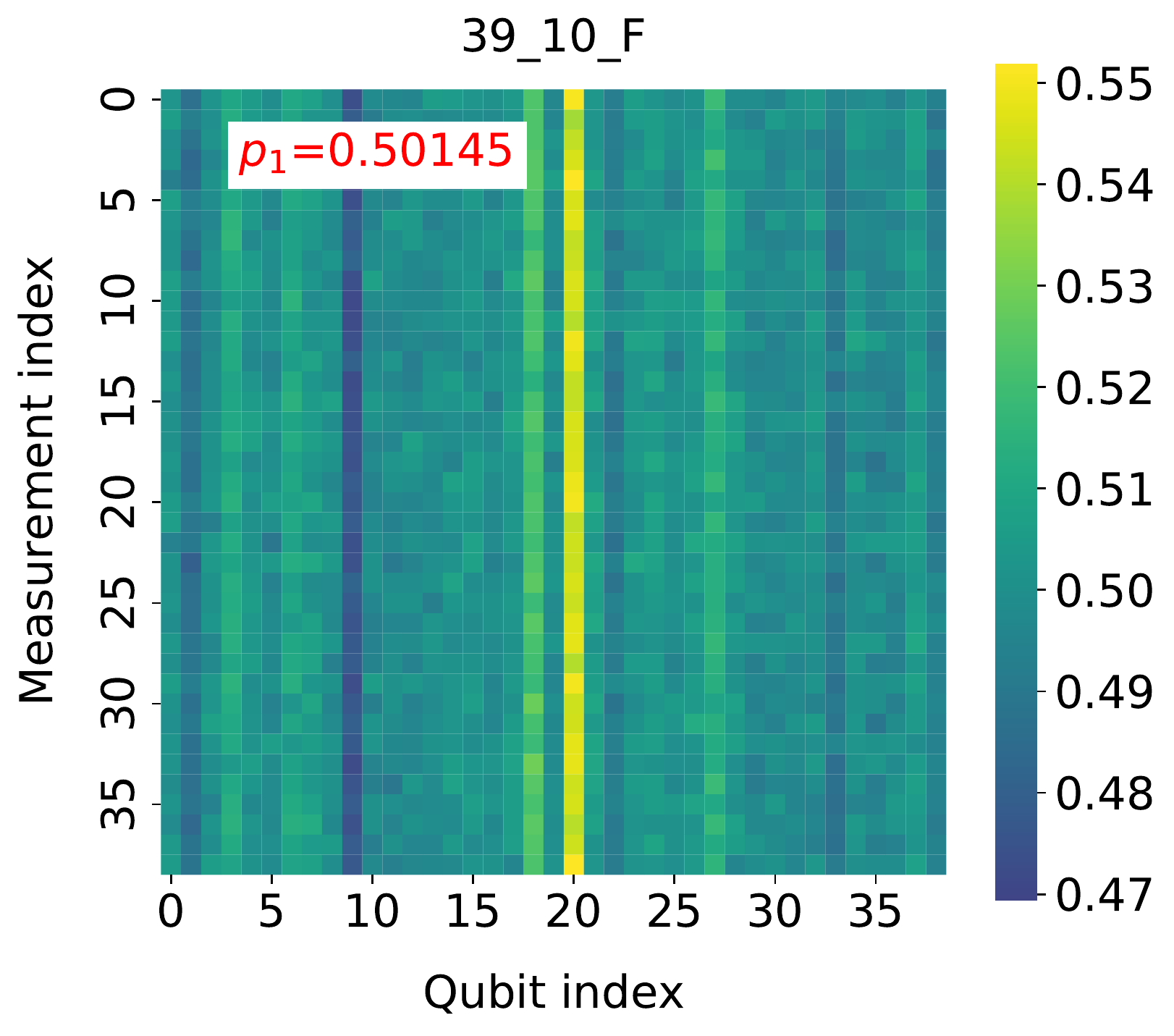}\\
\includegraphics[width=0.25\textwidth]{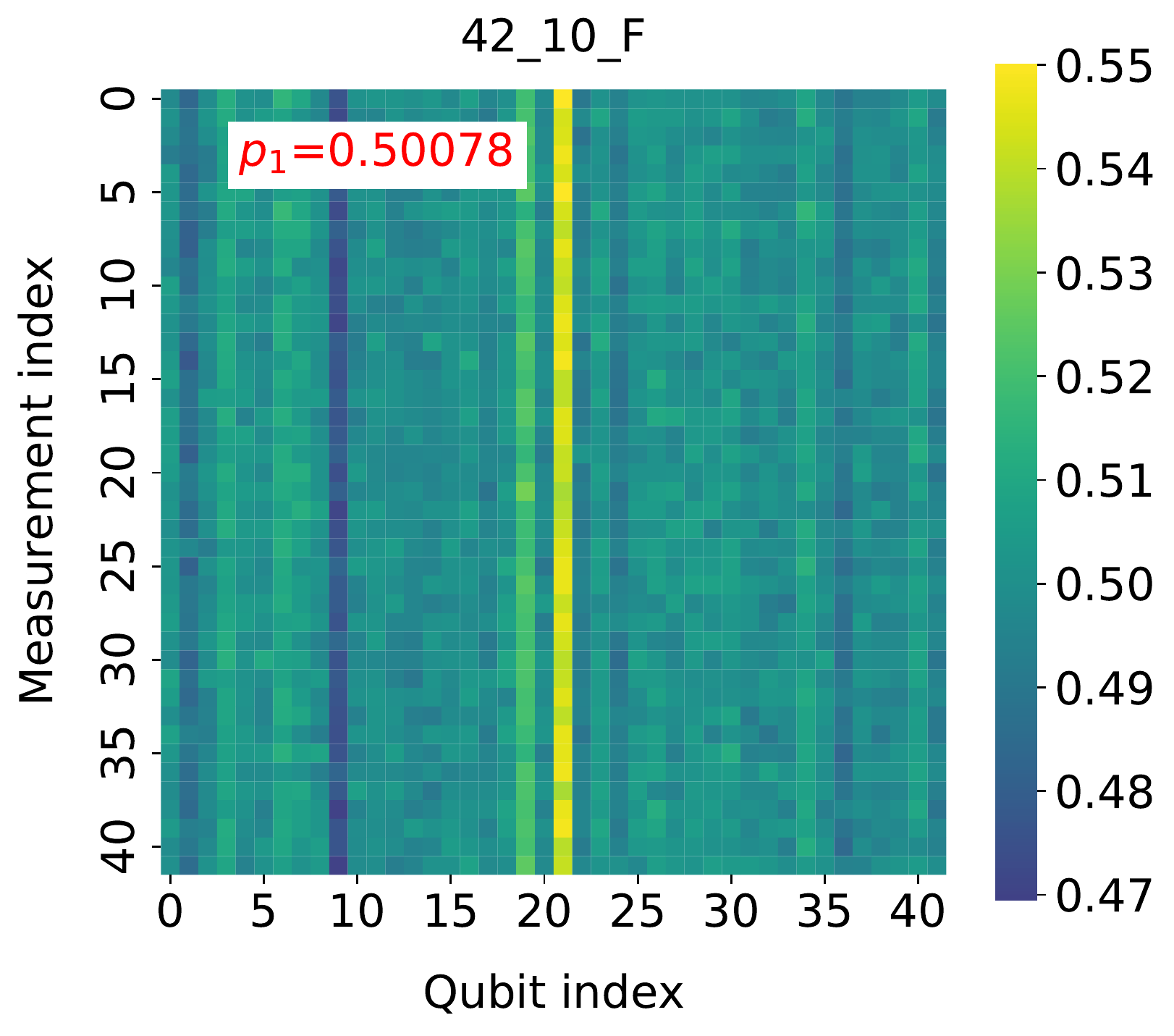}
\includegraphics[width=0.25\textwidth]{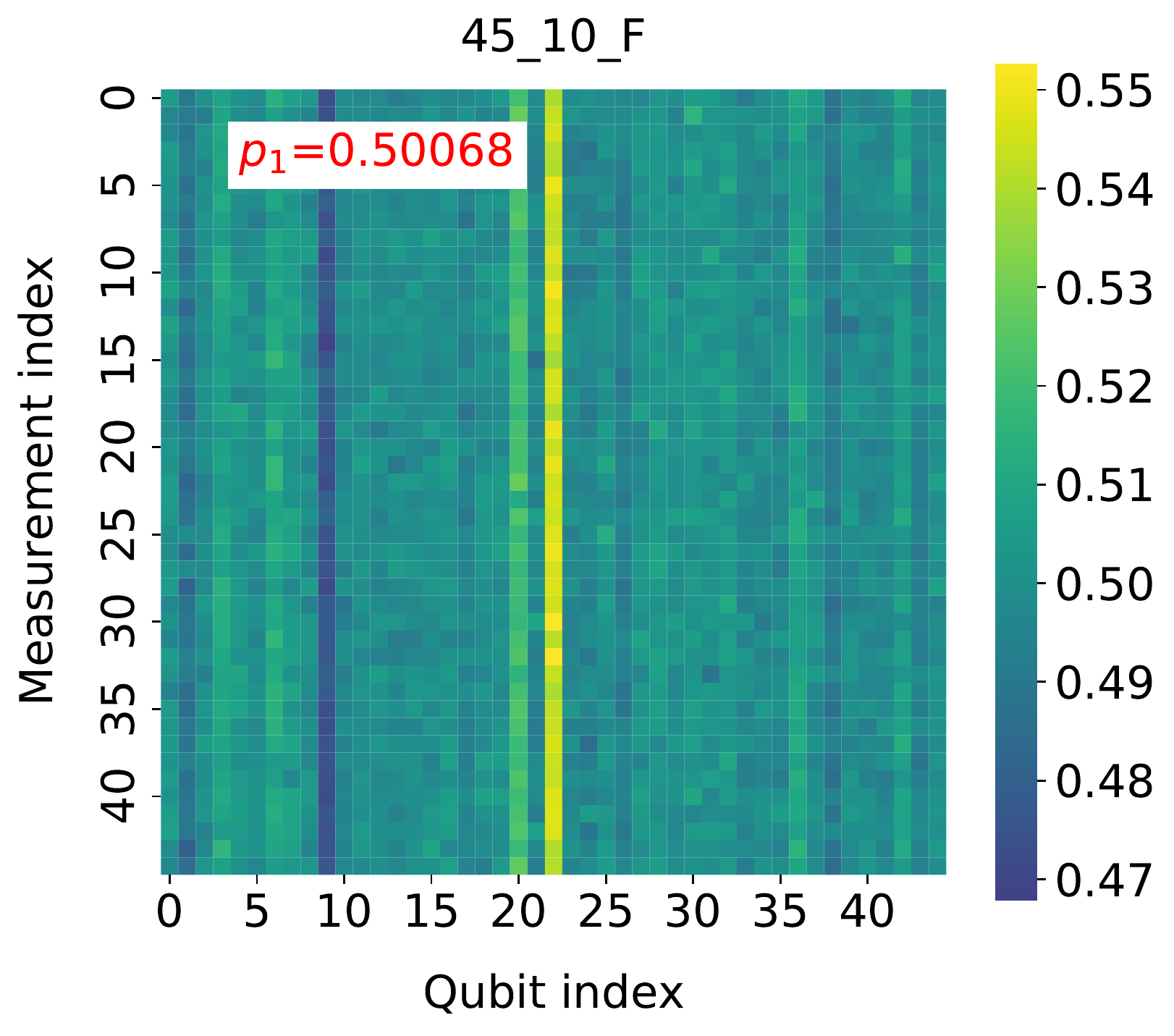}
\includegraphics[width=0.25\textwidth]{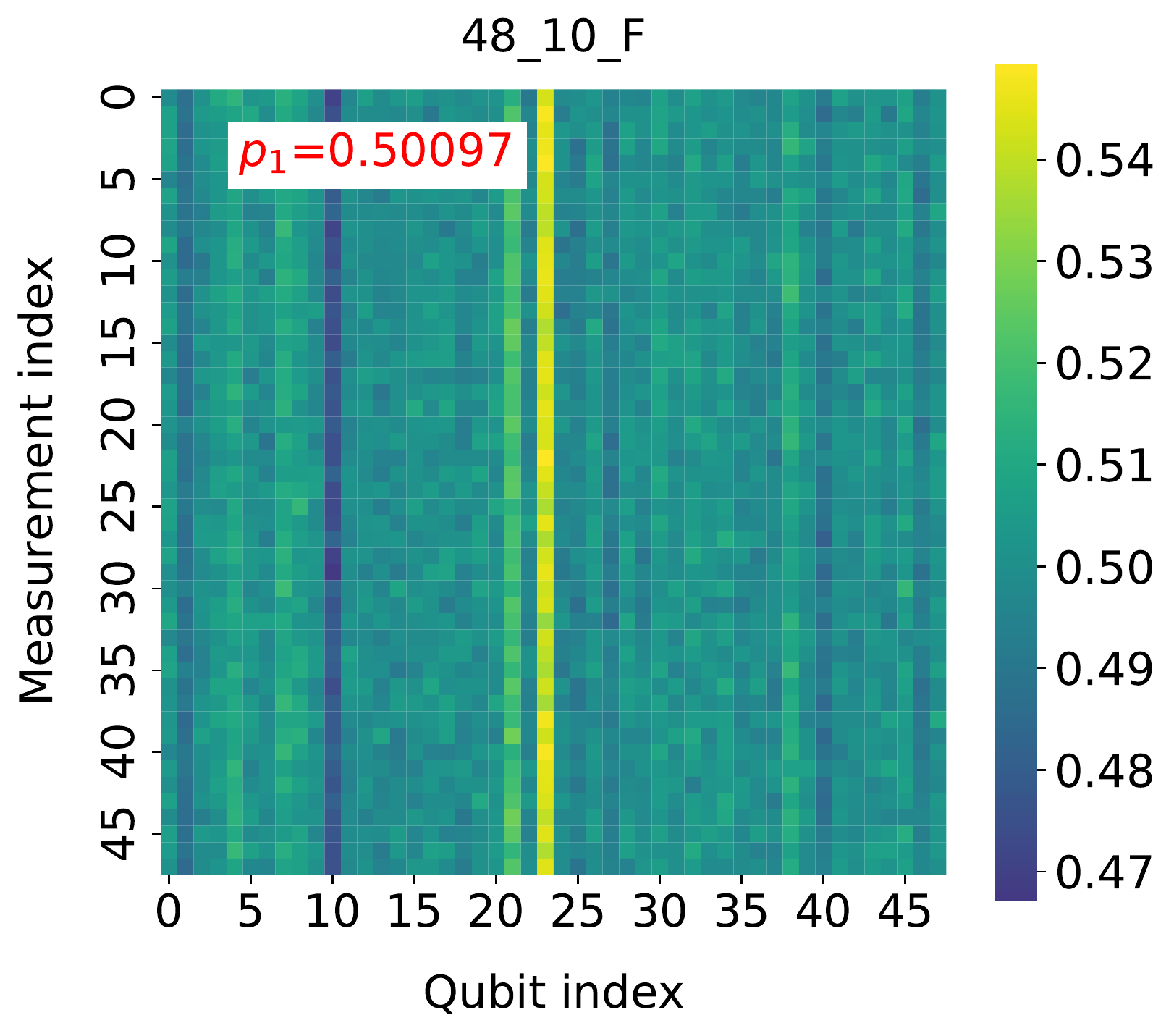}\\
\includegraphics[width=0.25\textwidth]{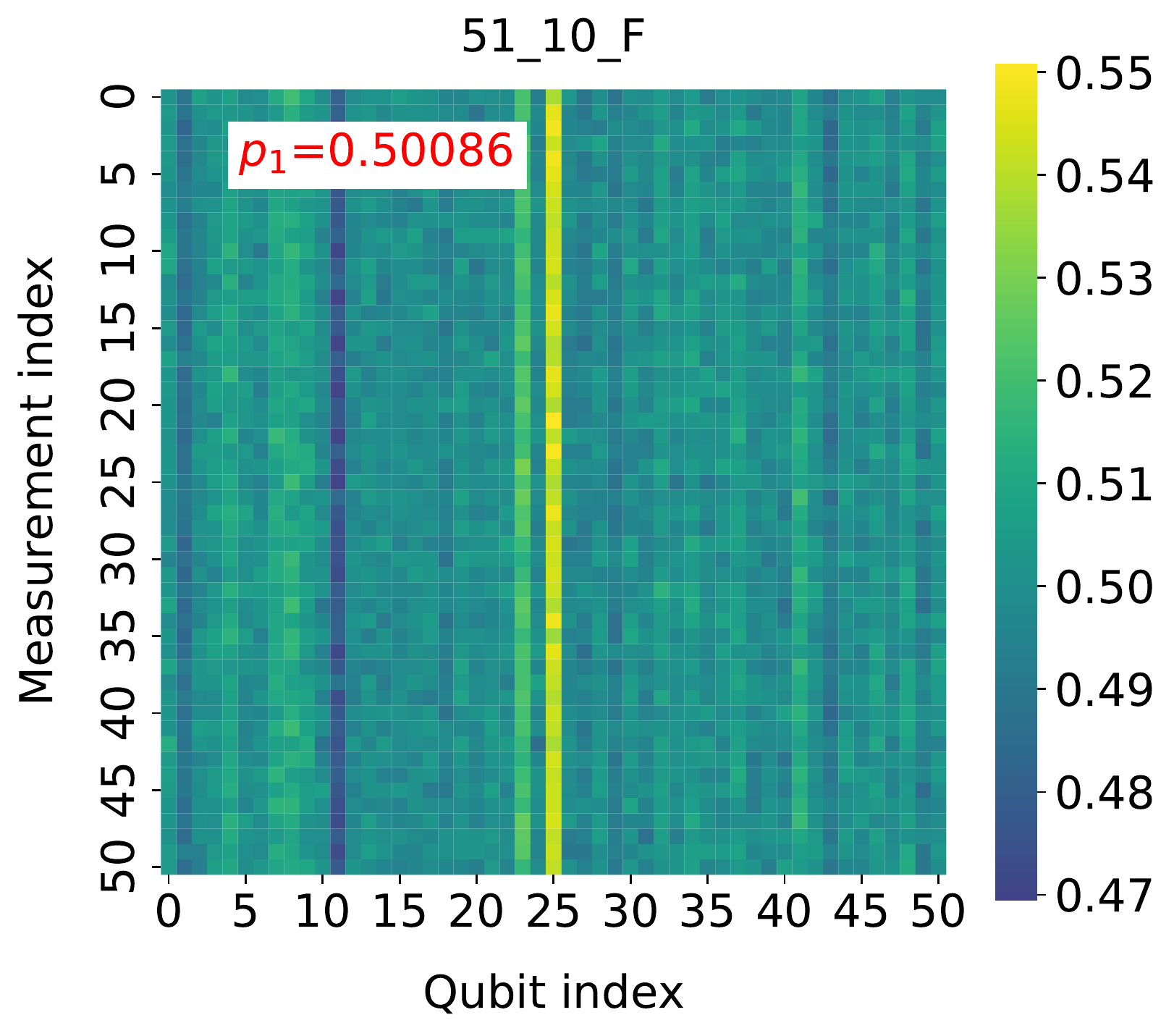}
\includegraphics[width=0.25\textwidth]{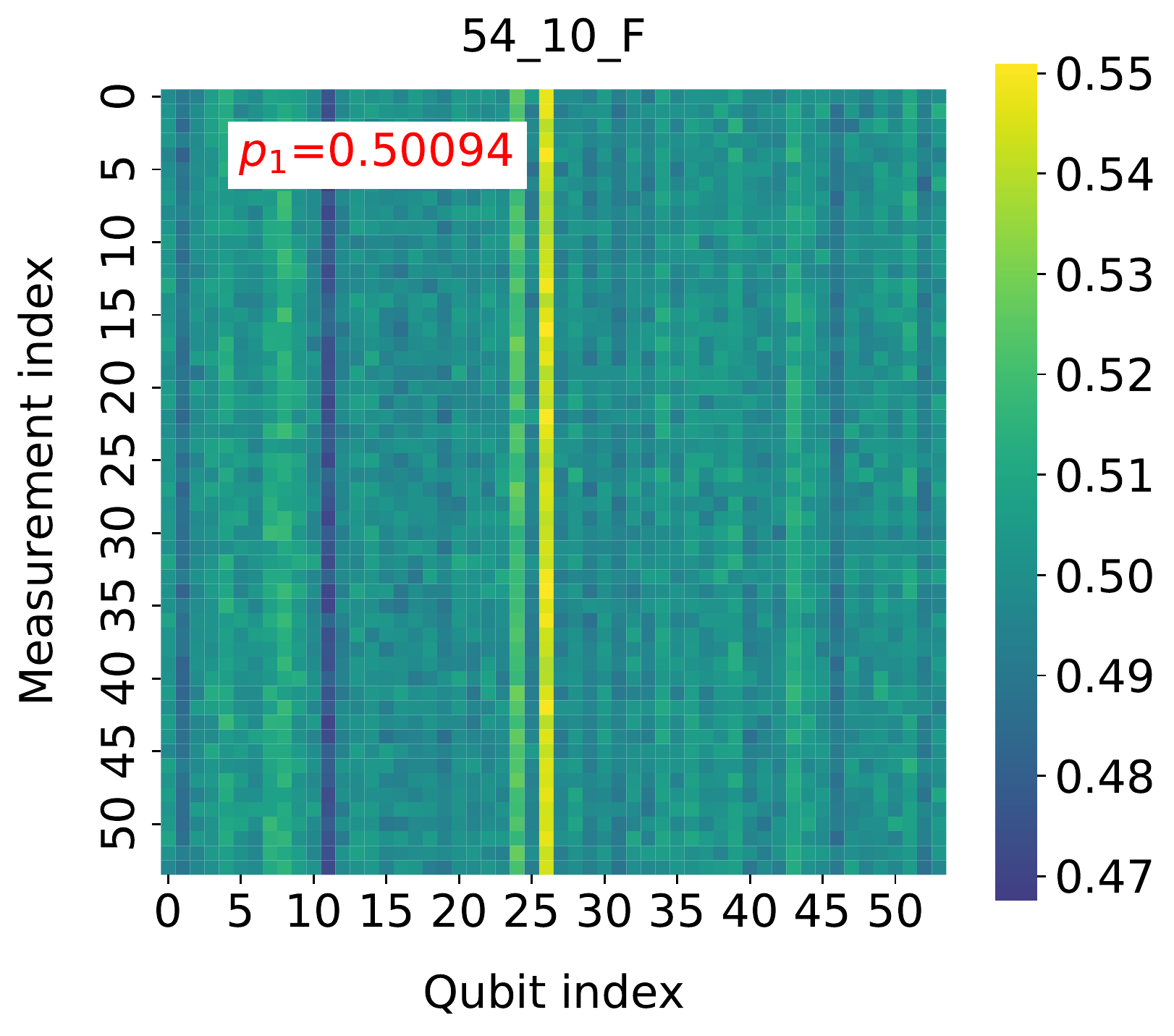}\hspace{0.25\textwidth}
\caption{Heatmaps of random bit-strings of Zuchongzhi with $n=15,18,21,...,51$ and $m=10$.}
\vspace{-10pt}
\end{figure*}
\vfill\pagebreak

\newpage
\begin{table*}[h]
\def\arraystretch{1.2}
\sffamily
\begin{tabular}{ m{3.0cm} r>{\RaggedRight\qquad} m{4.0cm} m{4.0cm}}
       \hline\hline\\[-5pt]
File name                     &$M$\phantom{xxx}       & \quad $p_1$           & NIST random number tests\\
\hline\hline
{\verb|15_10_F.csv|} &	 1000000 	 &0.5042914666666667   &Nonrandom \\
{\verb|18_10_F.csv|} &	 1000000 	 &0.5039382222222222   &Nonrandom \\
{\verb|21_10_F.csv|} & 	 1000000 	 &0.5016010476190477   &Nonrandom \\
\rowcolor{gray!20}
{\verb|24_10_F.csv|} & 	 1000000 	 &0.502242875          &Random \\
\rowcolor{gray!20}
{\verb|27_10_F.csv|} & 	 1000000 	 &0.5015492592592593   &Random\\
\rowcolor{gray!20}
{\verb|30_10_F.csv|} & 	 1000000 	 &0.5014525333333333   &Random\\
{\verb|33_10_F.csv|} & 	 1000000 	 &0.5014795757575757   &Nonrandom \\
{\verb|36_10_F.csv|} & 	 1000000 	 &0.5012370277777778   &Nonrandom \\
\rowcolor{gray!20}
{\verb|39_10_F.csv|} & 	 1000000 	 &0.5014506153846154   &Random \\
{\verb|42_10_F.csv|} & 	 1000000 	 &0.5007865238095238   &Nonrandom \\
\rowcolor{gray!20}
{\verb|45_10_F.csv|} & 	 1000000 	 &0.5006884222222222   &Random \\
\rowcolor{gray!20}
{\verb|48_10_F.csv|} & 	 1000000 	 &0.5009762291666666   &Random \\
{\verb|51_10_F.csv|} & 	 1000000 	 &0.5008632549019608   &Nonrandom \\
{\verb|54_10_F.csv|} &	 1050000 	 &0.5009445502645503   &Nonrandom \\
\rowcolor{gray!20}
{\verb|56_10_F.csv|} & 	 1170000 	 &0.5007047619047619   &Random \\ 
\hline\hline
\end{tabular}
\caption{NIST random number tests for random bit-strings of Zuchongzhi with $n=15,...,56$ and $m=10$.
$M$ stands for the number of bit-strings and $p_1$ for the probability of finding bit 1. 5 data files in directories,
\texttt{a1, a2, a3, a4, a5}, are merged into a single file. Nonrandom means that bit-strings fail some of the NIST random tests.}
\end{table*}
\vfill

\begin{table*}[ht]
\begin{center}
\begin{small}
\begin{verbatim}
Test Data File:/home/soh/Desktop/data/Zuchongzhi_nxx_m10/56_10_F_1M.csv

Type of Test                                        P-Value                 Conclusion
01. Frequency Test (Monobit)                        0.04725583993159622     Random
02. Frequency Test within a Block                   0.7377345740502045      Random
03. Run Test                                        0.9013653598338198      Random
04. Longest Run of Ones in a Block                  0.5576699582990581      Random
05. Binary Matrix Rank Test                         0.13573737861315063     Random
06. Discrete Fourier Transform (Spectral) Test      0.4912971242158931      Random
07. Non-Overlapping Template Matching Test          0.7861450901488796      Random
08. Overlapping Template Matching Test              0.3562201622350592      Random
09. Maurer's Universal Statistical test             0.18901932905865293     Random
10. Linear Complexity Test                          0.6012867491687948      Random
11. Serial test:
                                                    0.9287873362845678      Random
                                                    0.6473140941950127      Random
12. Approximate Entropy Test                        0.9658862333540649      Random
13. Cummulative Sums (Forward) Test                 0.08552956685396966     Random
14. Cummulative Sums (Reverse) Test                 0.09230410462727737     Random
15. Random Excursions Test:
                State       Chi Squared             P-Value                 Conclusion
                -4          2.2829376648618633      0.8087695808555937      Random
                -3          2.9604                  0.706091109222725       Random
                -2          5.783950617283951       0.3278123367522591      Random
                -1          6.75                    0.23990688227010742     Random
                +1          2.916666666666667       0.7128316836671243      Random
                +2          1.876543209876543       0.8659503247833513      Random
                +3          2.6755999999999998      0.7498477047163237      Random
                +4          4.022629459947245       0.5461624839081891      Random
16. Random Excursions Variant Test:
                State       COUNTS                  P-Value                 Conclusion
                -9.0        6                       0.5286121252556878      Random
                -8.0        9                       0.5761501220305789      Random
                -7.0        17                      0.7793054481711047      Random
                -6.0        21                      0.8961247779728454      Random
                -5.0        12                      0.5637028616507731      Random
                -4.0        11                      0.47819532084058625     Random
                -3.0        14                      0.5186050164287256      Random
                -2.0        17                      0.5596689271994115      Random
                -1.0        19                      0.47048642205878966     Random
                +1.0        29                      0.47048642205878966     Random
                +2.0        33                      0.4532547047537364      Random
                +3.0        49                      0.1065831695748876      Random
                +4.0        56                      0.08085559837005228     Random
                +5.0        60                      0.08326451666355045     Random
                +6.0        64                      0.08172275229865938     Random
                +7.0        63                      0.11846489387154538     Random
                +8.0        58                      0.20511768103340522     Random
                +9.0        39                      0.5995101785600032      Random
\end{verbatim} 
\end{small}
\end{center}
\caption{Result of the NIST random number tests for Zuchongzhi's data \texttt{56-10-F.csv}.}
\end{table*}

\begin{table*}[h]
\def\arraystretch{1.1}
\sffamily
\begin{tabular}{ m{3.0cm} r>{\RaggedRight\qquad} m{4.0cm} m{4.0cm}}
       \hline\hline\\[-5pt]
File name                     &$M$\phantom{xxx}       & \quad $p_1$           & NIST random number tests\\
\hline\\[-6pt]
{\verb|56_12_F_s11.csv|}      &1000000     &0.5004180892857143     &Nonrandom \\
\rowcolor{gray!20}
{\verb|56_12_F_s12.csv|}      &1000000     &0.4997393035714286     &Random\\
{\verb|56_12_F_s13.csv|}      &1000000     &0.5016109285714285     &Nonrandom \\
\rowcolor{gray!20}
{\verb|56_12_F_s15.csv|}      &1000000     &0.5005064107142857     &Random\\
\rowcolor{gray!20}
{\verb|56_12_F_s3.csv|}       &1000000     &0.5004619464285714     &Random\\
{\verb|56_12_F_s4.csv|}       &1000000     &0.5015308571428572     &Nonrandom \\
\rowcolor{gray!20}
\rowcolor{gray!20}
{\verb|56_12_F_s5.csv|}       &1000000     &0.5012426964285714     &Random\\
\rowcolor{gray!20}
{\verb|56_12_F_s6.csv|}       &1000000     &0.5019179107142857     &Random\\
\rowcolor{gray!20}
{\verb|56_12_F_s8.csv|}       &\; 980000     &0.5011627733236151   &Random\\
{\verb|56_12_F_s9.csv|}       &1000000     &0.5012499464285715     &Nonrandom \\[4pt]
{\verb|56_14_F_s11.csv|}      &1370000     &0.502104027632951     &Nonrandom\\
{\verb|56_14_F_s12.csv|}      &1440000     &0.5003947048611112    &Nonrandom\\
{\verb|56_14_F_s13.csv|}      &1700000     &0.5007865021008403    &Nonrandom\\
{\verb|56_14_F_s15.csv|}      &1120000     &0.5003659757653062    &Nonrandom\\
{\verb|56_14_F_s3.csv |}      &1100000     &0.5011199512987013    &Nonrandom\\
{\verb|56_14_F_s4.csv |}      &1290000     &0.5018366832779624    &Nonrandom\\
{\verb|56_14_F_s5.csv |}      &1310000     &0.5014550572519084    &Nonrandom\\
{\verb|56_14_F_s6.csv |}      &1400000     &0.5016173852040816    &Nonrandom\\
{\verb|56_14_F_s8.csv |}      &1480000     &0.5015874396718146    &Nonrandom\\
{\verb|56_14_F_s9.csv |}      &1390000     &0.5016998843782117    &Nonrandom\\[4pt]
{\verb|56_16_F_s11.csv|}      &3560000     &0.501555216693419     &Nonrandom\\
\rowcolor{gray!20}
{\verb|56_16_F_s12.csv|}      &3770000     &0.5006029840848807    &Random \\
{\verb|56_16_F_s13.csv|}      &3680000     &0.5012373495729814    &Nonrandom \\
{\verb|56_16_F_s15.csv|}      &2870000     &0.5008621329019413    &Nonrandom \\
{\verb|56_16_F_s3.csv|}       &2890000     &0.5017474913494809    &Nonrandom \\
{\verb|56_16_F_s4.csv|}       &3200000     &0.5013628404017857    &Nonrandom \\
{\verb|56_16_F_s5.csv|}       &2920000     &0.5025822773972602    &Nonrandom \\
{\verb|56_16_F_s6.csv|}       &3570000     &0.5023631752701081    &Nonrandom \\
{\verb|56_16_F_s8.csv|}       &3710000     &0.5017546592221794    &Nonrandom \\
{\verb|56_16_F_s9.csv|}       &3420000     &0.5018155127401838    &Nonrandom \\[4pt]
{\verb|56_18_F_s11.csv|}     &10080000     &0.5012828125          &Nonrandom \\
\rowcolor{gray!20}
{\verb|56_18_F_s12.csv|}     &9730000      &0.5006819630010277    &Random    \\
{\verb|56_18_F_s13.csv|}     &11530000     &0.5014127137281625    & Nonrandom \\
\rowcolor{gray!20}
{\verb|56_18_F_s15.csv|}     & 9530000     &0.5014926154249738    & Random    \\
{\verb|56_18_F_s3.csv|}      & 7660000     &0.5021444307161507    & Nonrandom \\
{\verb|56_18_F_s4.csv|}      & 7510000     &0.5024223606619745    & Nonrandom \\
{\verb|56_18_F_s5.csv|}      & 7960000     &0.5020297963029433    & Nonrandom \\
{\verb|56_18_F_s6.csv|}      & 9390000     &0.5027202723261829    & Nonrandom \\
{\verb|56_18_F_s8.csv|}      & 8620000     &0.501850381173351     & Nonrandom \\
{\verb|56_18_F_s9.csv|}      & 7470000     &0.5023868091413272    & Nonrandom \\[4pt]
{\verb|56_20_F_s11.csv|}     &27900000     &0.5018168573988735    & Nonrandom \\
{\verb|56_20_F_s12.csv|}     &25870000     &0.5013624744602132    & Nonrandom \\
{\verb|56_20_F_s13.csv|}     &26580000     &0.5017120135977642    & Nonrandom \\
{\verb|56_20_F_s15.csv|}     &20100000     &0.5013940840440654    & Nonrandom \\
{\verb|56_20_F_s3.csv|}      &19100000     &0.502459691473448     & Nonrandom \\
{\verb|56_20_F_s4.csv|}      &21510000 	   &0.5024990527661553    & Nonrandom \\
{\verb|56_20_F_s5.csv|} 	 &22570000 	   &0.5026209222102664    & Nonrandom \\
{\verb|56_20_F_s6.csv|} 	 &24510000 	   &0.5030497945445008    & Nonrandom \\
{\verb|56_20_F_s8.csv|} 	 &19840000 	   &0.5022362093173963    & Nonrandom \\
{\verb|56_20_F_s9.csv|} 	 &23510000 	   &0.5026191111684998    & Nonrandom \\
\hline\hline
\end{tabular}
\caption{NIST random number tests for random bit-strings of Zuchongzhi with $n=56$ and $m=12,14,16,18,20$.
$M$ stands for the number of bit-strings and $p_1$ for the probability of finding bit 1. 
Nonrandom means that bit-strings fail some of the NIST random tests.}
\end{table*}
\end{widetext}

\begin{table*}[t]
\def\arraystretch{1.5}
\begin{tabular} {l|m{2cm}m{2cm}m{2cm} m{2cm}m{2cm}m{2cm}}
\hline \hline
              &classical n53 &n53-m12             & n53-m14            & n53-m16              & n53-m18            & n53-m20  \\
\hline
classical n53 &0.0        &0.02495338 &0.02459099  &0.02560033 &0.02514710 &0.0249869\\
n53-m12       &0.02495337 &0.0        &0.00083951  &0.00040303 &0.00050475 &0.0005789\\
n53-m14       &0.02459099 &0.00083951 &0.0         &0.00099459 &0.00128885 &0.0013602\\
n53-m16       &0.02560033 &0.00040303 &0.00099459  &0.0        &0.00066794 &0.0007389\\
n53-m18       &0.02514710 &0.00050475 &0.00128885  &0.00066794 &0.0        &0.0002826\\
n53-m20       &0.02498690 &0.00057892 &0.00136027  &0.00073890 &0.00028265 &0.0  \\
\hline
\hline
\end{tabular}
\caption{For $n=53$, the Wasserstein distances among classical random bit-strings and Sycamore random bit-strings with cycle 
$m=12,14,16,18,20$. The Wasserstein distance between the classical random bit-strings and Sycamore's random bit-strings 
is much larger than those between Sycamore's random bit-strings with different cycle.}
\end{table*}

\begin{table*}[t]
\def\arraystretch{1.5}
\begin{tabular} {l|m{2cm}m{2cm}m{2cm} m{2cm}m{2cm}m{2cm}}
\hline \hline
              &classical n56  &n56-m12     &n56-m14     &n56-m16     &n56-m18     &n56-m20\\ \hline
classical n56 &0.0            &0.00173514  &0.00183101  &0.00171989  &0.00161961  &0.00174550\\
n56-m12       &0.00173514     &0.0         &0.00071221  &0.00023922  &0.00056779  &0.00034071\\
n56-m14       &0.00183101     &0.00071221  &0.0         &0.00085164  &0.00125725  &0.00073764\\
n56-m16       &0.00171989     &0.00023922  &0.00085164  &0.0         &0.00045474  &0.00025549\\
n56-m18       &0.00161961     &0.00056779  &0.00125725  &0.00045474  &0.0         &0.00056125\\
n56-m20       &0.00174550     &0.00034071  &0.00073764  & 0.00025549 &0.00056125  &0.0  \\
\hline
\hline
\end{tabular}
\caption{For $n=56$, the Wasserstein distances among classical random bit-strings and Zuchongzhi random bit-strings with cycle
$m=12,14,16,18,20$.}
\end{table*}

\end{document}